\documentclass[prd,aps,twocolumn,showpacs,amsmath,amssymb,amsfonts]{revtex4}

\usepackage{graphicx}

\usepackage[dvips,usenames]{color}
\usepackage{dsfont}
\usepackage{hyperref}
\usepackage{bm}

\def\edth{\;\raise1.0pt\hbox{$'$}\hskip-6pt\partial}
\def\baredth{\;\overline{\raise1.0pt\hbox{$'$}\hskip-6pt
\partial}}

\newcommand{\lss}{{\textrm{LSS}}}
\newcommand{\etaJP}{{\alpha_\ell}}
\newcommand{\etaJPdemi}{{\alpha_{L}}}
\newcommand{\tildeetaJPdemi}{{\hat{\alpha}_{L}}}
\newcommand{\etaJPsqrt}{{\alpha_{\tilde L}}}
\newcommand{\bk}{{\bm k}}

\newcommand{\vk}{{\bm k}}

\newcommand{\vx}{{\bm x}}

\newcommand{\dd}{{\rm d}}
\newcommand{\ii}{{\rm i}}
\newcommand{\mG}{{\cal G}}

\newcommand{\mO}{{\cal O}}

\newcommand{\mR}{{\cal R}}

\newcommand{\hn}{{\bm \hat{n}}}

\newcommand{\hk}{{\bm \hat{k}}}

\newcommand\spart{\;\raise1.0pt\hbox{$/$}\hskip-6pt\partial}
\newcommand\spartb{\;\overline{\raise1.0pt\hbox{$/$}\hskip-6pt
\partial}}
\newcommand{\wtroisj}[6]{\left( 
\begin{array}{ccc}
        \! #1\! & #2\!  & #3\!  \\
         \! #4\! & #5\!  & #6\!         
\end{array}\right)}

\newcommand{\vn}{{\bm n}}
\newcommand{\jj}{j}

\begin{document}
\title{CMB spectra and bispectra calculations: making the flat-sky approximation rigorous}

\author{Francis Bernardeau$^1$} 
\email{francis.bernardeau@cea.fr}
\author{Cyril Pitrou$^2$}
\email{cyril.pitrou@port.ac.uk}
\author{Jean-Philippe Uzan$^{3,4,5}$}
\email{uzan@iap.fr}
\affiliation{$^1$ Institut de Physique Th\'eorique,
 CEA, IPhT, F-91191 Gif-sur-Yvette, France
 CNRS, URA 2306, F-91191 Gif-sur-Yvette, France\\
$^2$Institute of Cosmology and Gravitation, Dennis Sciama Building, Burnaby Road, 
                        Portsmouth, PO1 3FX, United Kingdom,\\
$^3$Institut d'Astrophysique de Paris, UMR-7095 du CNRS, 
                       Universit\'e Pierre et Marie Curie, 98 bis bd Arago, 75014 Paris, France\\
$^4$Department of Mathematics and Applied Mathematics, Cape Town University, Rondebosch 7701, South Africa\\
$^5$National Institute for Theoretical Physics (NITheP), Stellenbosch 7600, South Africa.}

\vskip 0.15cm

\date{\today}
\pacs{98.80.-k}

\begin{abstract}
This article constructs flat-sky approximations in a controlled way in the context of the cosmic microwave background observations for
the computation of both spectra and bispectra. For angular spectra, it is explicitly shown that there exists a whole family of
flat-sky approximations of similar accuracy for which the expression and amplitude of next to leading order terms can be explicitly
computed. It is noted that in this context two limiting cases can be encountered for which the expressions can be further 
simplified. They correspond to cases where either the sources are localized in a narrow region (thin-shell approximation)
or are slowly varying over a large  distance (which leads to the so-called Limber approximation).

Applying this to the calculation of the spectra it is shown that, as
long as the late integrated Sachs-Wolfe 
contribution is neglected, the flat-sky approximation at leading order is accurate 
at 1\% level for any multipole. 

Generalization of this construction scheme to the bispectra led to the introduction of an alternative description of the
bispectra for which the flat-sky approximation is well controlled. This is not the case for the usual description of the bispectrum
in terms of reduced bispectrum for which a flat-sky approximation is proposed but the next-to-leading order
terms of which remain obscure.  
\end{abstract}
\maketitle

\section{Introduction}

Cosmological surveys in general and the cosmic microwave background (CMB) in particular 
are naturally constructed on our celestial sphere. Because of the statistical isotropy of such observations,
cosmological statistical properties, such as the angular power spectra or the bispectra, are better captured in reciprocal
space, that is in harmonic space. In general however, most of the physical mechanisms at play take place at small scale
and are therefore not expected to affect the whole sky properties. For instance, the physics of the CMB is, to a large extent, 
determined by sub-Hubble interactions corrsponding to sub-degree scale on our observed sky. Decomposition in spherical 
harmonics, while introducing a lot of complication of the calculations, does not carry much physical insight into these
mechanisms so rather blurs the physics at play.

In this respect, a flat-sky approximation, in which the sky is approximated by a 2-dimensional plane tangential
to the celestial sphere, hence allowing the use of simple Cartesian
Fourier transforms, drastically simplifies CMB computations.  Such an approximation is intuitively expected to be accurate at small scales. 
So far this approximation is mostly based on an heuristic correspondence between the two sets of harmonic basis (spherical
and Euclidean) which can be summarized for a scalar valued observable 
by~\cite{1997PhRvD..55.1830Z,2000PhRvD..62d3007H}
\begin{equation}\label{corres}
\Theta(\vn) = \sum a^{\Theta}_{\ell m}Y_\ell^m(\hat \vn)\to \int \dd^2 {\bf l}\Theta({\bf l})e^{\ii {\bf l}.{\bf \theta}}\,.
\end{equation}
In the context of CMB computation, the relations between the flat-sky and the full-sky expansions 
have been obtained at leading order in Ref.~\cite{2000PhRvD..62d3007H}. 
However, its validity for the angular power spectrum  and the bispectrum is not yet understood in 
the general case and  the expression and order of magnitude of the next to leading order terms are 
still to be computed. 

The goal of this article is to provide such a systematic construction. In particular, we will show that
there exists a two-parameter family of flat-sky approximations for which well-controlled expansions can be built.
That allows us to discuss in details their accuracy by performing the computation up to next-to-leading
order. In the new route we propose here,  we derive the flat-sky expansion directly on the 2-point angular 
correlation function, instead of relying on the correspondence~(\ref{corres}). One of
the technical key step is to expand the eigenfunctions of the spherical Laplacian 
onto the eigenfunctions of the cylindrical Laplacian in order to relate
the (true) spherical coordinates on the sky expansion to the cylindrical coordinates of the flat-sky expansion.
This approach proves very powerful since it enables to obtain the full series of corrective terms 
to the flat-sky expansion. 


Once the method has been developed, it can be generalized to the polarisation
and also to the computation of the bispectrum. In this latter case, depending on the way
one chooses to describe the bispectrum, the exact form
of the corrective terms has not been obtained but we can still provide
an approximation whose validity can be checked numerically.

Before we enter the details of our investigations, and as the literature can be very confusing 
regarding the flat-sky approximations, let us present the different levels of approximations we are going to use.
The reason there exist at all a flat-sky approximation is that the physical processes at play have a finite 
angular range. In case  of the CMB, most of physical processes take place at sub-horizon scales and within the 
last scattering surface (LSS) (to the exception of the late integrated Sachs-Wolfe effect) and therefore
within 1 degree scale on the sky.
Let us denote $\ell_{0}$ the scale,  in harmonic space associated with this angular scale. While using the flat 
sky approximation, the physical processes will be computed in a slightly deformed geometry (changing a conical region 
into a cylindrical one) introducing a priori an error
of the order of $1/\ell_{0}$  (actually in $1/\ell_{0}^2$ depending on the type of source terms as it
will be discussed in details below). Another part of the approximation 
is related to the projection effects which determine the link between physical quantities and observables. It introduces another
layer of approximation of purely geometrical origin. For that part the errors behave a priori as $1/\ell$ if $\ell$  is the scale
of observation in harmonic space. 

The resulting integrals do not lead to factorizable properties as 
it is the case for exact computations, while a factorization property can be recovered taking advantage
of two possible limiting situations. First, for most of the small
scale physical processes, one can use the fact that the radial
extension of the source is much smaller than its distance from the observer. 
It is then possible to perform a {\em thin-shell approximation} effectively assuming
that all sources are at the same distance from the observer.

Another limit case corresponds to the situation in which the source
terms are slowly varying and spread over a wide range of distances, as e.g. for galaxy distribution or weak-lensing field. In this case
the sources support appears very elongated and it is then possible to use the
{\em Limber approximation}~\cite{1953ApJ...117..134L,1980lssu.book.....P,2008PhRvD..78l3506L} 
which takes advantage of the fact that
contributing wave modes in the radial direction should be much smaller that the modes in the transverse direction
(but as such the Limber approximation can be used in conjunction of the flat-sky approximation or not).  

These different layers of approximations proved to be useful to compute efficiently the effects of secondaries 
such as lensing, but also of great help for computing the effects of non-linearities 
at the LSS contributing to the bispectrum, either analytically~\cite{2009JCAP...08..029B,Pitrou:2008ak} or numerically~\cite{2010JCAP...07..003P} (see Ref.~\cite{2009JCAP...05..014N} 
to compare to the full-sky expressions) as well as for the angular
power spectrum; see 
e.g. Ref.~\cite{2006PhR...429....1L} for a review and for the relation between the flat-sky 
and the full-sky expansions in both real and harmonic space. We shall thus
detail the expressions and corrective terms of the flat-sky approximation in these
two approximations. In particular, we recover the result by Ref.~\cite{2008PhRvD..78l3506L} with a different method
in the case of the Limber approximation. This is a consistency check of our new method.\\



First, we consider the computation of the angular power spectrum in Section~\ref{SecSpectrumScalaire}
starting with an example of such a construction in order to show explicitly how to construct
next to leading order terms whose correction is found to be of the order of $1/\ell^2$. 
We then show that this construction is not unique and present the construction of a whole family of 
approximations whose relationship can be explicitly uncovered. 
In Section~\ref{SecApproximations} we present further computation approximations, e.g. 
the Limber (\S~\ref{limbersec}) and thin-shell (\S~\ref{thinsec}) approximations. 
While in Sections~\ref{SecSpectrumScalaire} and ~\ref{SecApproximations} we have assumed, for clarity but also
because it changes the result only at next-to-leading order, that the transfer function
was scalar, in Section~\ref{SecSpectrumGeneral} we provide the general case of the flat-sky approximation 
up to next-to-leading order corrections in $1/\ell$ including all physical effects. 
Eventually Section~\ref{SecPolar} considers the case of higher spin quantities to describe the CMB polarization.

We explore the case of the bispectrum construction in Section~\ref{SecBispectre}. 
One issue we encountered here is that different equivalent parameterizations can be used to describe bispectra 
(amplitudes of bispectra depend on both the scale and shape of the triangle formed by three $\ell$ modes
that can be described in different manners). 
We thus present an alternative description of the bispectrum for which the flat-sky approximation can
be done in a controlled way. Although we did not do the calculation explicitly, next-to-leading order terms
can be then obtained. This is not the case for the reduced bispectrum for which
we could nonetheless propose a general flat-sky approximation. Similarly to the case of spectra
practical computations can then be done in the thin-shell approximation or
the Limber approximation.



\section{Power spectrum in the flat-sky limit}\label{SecSpectrumScalaire}

\subsection{General definitions}

In the line of sight approach, the temperature $\Theta(\hn)$ observed in a direction $\hn$ is
expressed as the sum of all emitting sources along the line of sight
in direction $\hn$,
\begin{equation}\label{Theta}
\Theta(\hn)=\int\dd r\int \frac{\dd^3 \vk}{(2 \pi)^{3/2}}  \,w(\bk,\hn,r) \Phi (\vk)\exp(\ii \vk.\vx)\,,
\end{equation}
where $\vx$ is the position at distance $r$ and angular position $(\theta,\varphi)$ and
$\Phi$ is the primordial gravitational potential from which all initial conditions can be constructed~\cite{1995ApJ...455....7M}.
$w(\bk, \hn,r)$ is a transfer function that depends on both the wave-number 
$\bk$ and the direction of observation $\hn$. In particular, it incorporates altogether the visibility function, $\tau' e^{-\tau}$, 
where $\tau$ is the optical depth, and the time and momentum dependencies of the sources. 
It can always be expanded as
\begin{equation}\label{Expsources}
w(\vk,\hat \vn,r) = \sum_{\jj,m} w_{\jj m}(k,r) (\ii)^\jj \sqrt{\frac{4 \pi}{2 \jj+1}} Y_{\vk}^{\jj m}(\hat \vn)
\end{equation}
where the $Y_{\vk}^{\jj m}$ are the spherical harmonics with azimutal direction 
aligned with $\vk$. The source multipoles are defined using the
same conventions as in Refs.~\cite{1997PhRvD..56..596H,2009CQGra..26f5006P}
except that the multipoles are defined here using the direction of
observation whereas in these references it is defined with the direction of
propagation~\footnote{The direction of observation is opposite to the
  direction of propagation, and the transformation properties under
parity bring an extra $(-1)^\ell$ in the above formula when compared to
Eq.~(7.1) of Ref.~\cite{2009CQGra..26f5006P} or Eq.~(10) of
Ref.~\cite{1997PhRvD..56..596H}.}. As long as we consider only scalar
type perturbations in the perturbation theory, the source term will only contain $w_{\jj m}$ terms with $m=0$. For instance, the Doppler term of the scalar
perturbation introduces a term $w_{10}$ etc.

In order to focus our attention to the geometrical properties of the flat-sky expansion, 
we first assume for simplicity that the temperature fluctuations do
not depend on $\hn$ and are thus only scalar valued functions. The statistical isotropy of the primordial fluctuations
then implies that the $\vk$-dependency reduces to a $k$-dependency,
so that the transfer function is of the form $w(k,r)=w_{00}$. The general case is
postponed to Section~\ref{SecSpectrumGeneral}.

The two-point angular correlation function of the temperature anisotropies, defined by
\begin{equation}\label{defxi}
\xi(\theta)=\langle \Theta(\hn) \Theta(\hn') \rangle_{\hn.\hn'=\cos \theta}\,,
\end{equation}
is related to the angular power spectrum $C_\ell$ by
\begin{equation}\label{inttheta}
C_{\ell}=2\pi \int\sin\theta\dd\theta\,P_{\ell}(\cos\theta)\,\xi(\theta),
\end{equation}
where $P_{\ell}$ are the Legendre polynomials of order $\ell$.
The 3-dimensional power spectrum of the primordial potential being defined by
\begin{equation}
\langle \Phi(\vk) \Phi(\vk') \rangle = \delta(\vk+\vk')P(k)\,,
\end{equation}
we can easily invert this relation to get a one-parameter family
of correlation functions $\xi_v(\theta)$ as
\begin{eqnarray}\label{xiexp2}
\qquad \xi_v(\theta)&=& \int \frac{\dd k_{r}}{(2\pi)^2}\ k_{\perp}\dd k_{\perp}  w(k,r) w^\star(k,r')\dd r \dd r'  \nonumber\\
& \times& P(k) J_{0}\left[k_{\perp}\left(r'\sin v \theta +r\sin (1-v)\theta \right)\right]\nonumber\\
&\times& \exp\left[\ii k_{r}\left(r \cos (1-v)\theta -r' \cos v\theta \right) \right]\,
\end{eqnarray}
where $J_{0}$ is the Bessel function of the first kind of order $0$
and where the star denotes the complex conjugation.
The index $v$ refers to the parametrization according to the two line-of-sight
\begin{eqnarray}\label{FSparameterizationJP}
\vx&=&r[\sin((1-v)\theta) \vn_\perp,\cos((1-v)\theta)]\\
\vx'&=&r'[\sin(v\theta) \vn'_\perp,\cos(v\theta)]\,,\nonumber
\end{eqnarray}
where we have defined the two-dimensional vectors $\vn_\perp\equiv(\cos \varphi,\sin \varphi)$ and
$\vn'_\perp\equiv(\cos (\varphi+\psi),\sin (\varphi+\psi))$ with $\psi=\pi$ so that $\vn'_{\perp}=-\vn_{\perp}$. 
Note that the relation (\ref{xiexp2}) has been obtained for a fixed value of $\psi$, the relative angle between $\vn'_{\perp}$ and $\vn_{\perp}$, although it could have been left as a free parameter. As we will see in the following, flat-sky approximations can 
indeed be obtained for any fixed value of $v$ and $\psi$ provided $\hn$ and $\hn'$ are close enough to the azimuthal direction. 
In the last part of \ref{SecRapConv} we will briefly comment on the
effect of considering $\psi$.

Any Fourier mode  $\vk$ can then be decomposed into a component $k_r$ orthogonal to $\vn_\perp$ and 
a component $\vk_\perp$ parallel to $\vn_\perp$. Its modulus
$k$ is thus to be  considered as a function of $k_r$ and $k_\perp $ since
\begin{equation}
k = \sqrt{k_r^2 + k_\perp^2}\,.
\end{equation}
To finish, we parameterize $\vk_\perp$ as
\begin{equation}\label{paramk}
\vk = (k_\perp \cos \beta,k_\perp \sin \beta,k_r)\,.
\end{equation}
Note that the Bessel function in the expression~(\ref{xiexp2}) arises from the integration
over the angle between $\vk_{\perp}$ and $\vn_{\perp}$, i.e. $\beta - \varphi$. This requires to 
assume that the transfer function depends neither on $\hat \vk=\vk/k$ nor on $\hn$ and 
is thus independent of the angle between $\vk_{\perp}$ and $\vn_{\perp}$.
Note that for simplicity we could have chosen to set $\varphi=0$ since only the relative angle 
between $\vk_\perp$ and $\vn_\perp$ matters in the derivation of the result.

\subsection{A construction case: the $v=0$ case}\label{SecIntuitiveComputation}

Before investigating the full family $\xi_v$ in the flat-sky
limit, let us concentrate on the particular case $v=0$. 
The flat-sky approximation is obtained as a small angle limit, i.e.
$\theta\ll1$ while letting $\etaJP=\ell\theta$ fixed. In order to
expand the Legendre polynomials $P_{\ell}(\cos\theta)$ in that
limit, we start from their integral representation as
\begin{equation}
P_{\ell}(\cos\theta)=\frac{1}{\pi}\int_{0}^{\pi}\exp\left[\ell\log\left(\cos\theta+\ii\sin\theta\cos\varphi\right)\right]\dd\varphi.
\end{equation}
In the above mentioned limit, it gives the integral representation of $J_{0}(\ell\theta)$,
\begin{equation}
J_{0}(\ell\theta)=\frac{1}{\pi}\int_{0}^{\pi}\exp\left[\ii \ell\theta\cos\varphi\right]\dd\varphi\,.
\end{equation}
Furthermore, it allows to obtain the subsequent terms of the expansion as
\begin{eqnarray}\label{Plexpan}
P_{\ell}(\cos\theta)&=&J_{0}({\etaJP})\\
&-&\frac{\theta}{2}J_{1}({\etaJP})-\frac{\theta^2}{24}J_{0}({\etaJP})+\frac{\theta^2}{12\,{\etaJP}}J_{1}({\etaJP})+\dots,\nonumber
\end{eqnarray}
again for a fixed ${\etaJP} \equiv \ell\theta$. The existence of such an expansion, and its simplicity, is
central in the construction we present here. 
It shows that the eigenfunctions of the Laplacian on the 2-sphere converge toward 
eigenfunctions of the Laplacian on the Euclidean plane. 
%

\begin{figure*}[!htb]
\includegraphics[width=8.5cm]{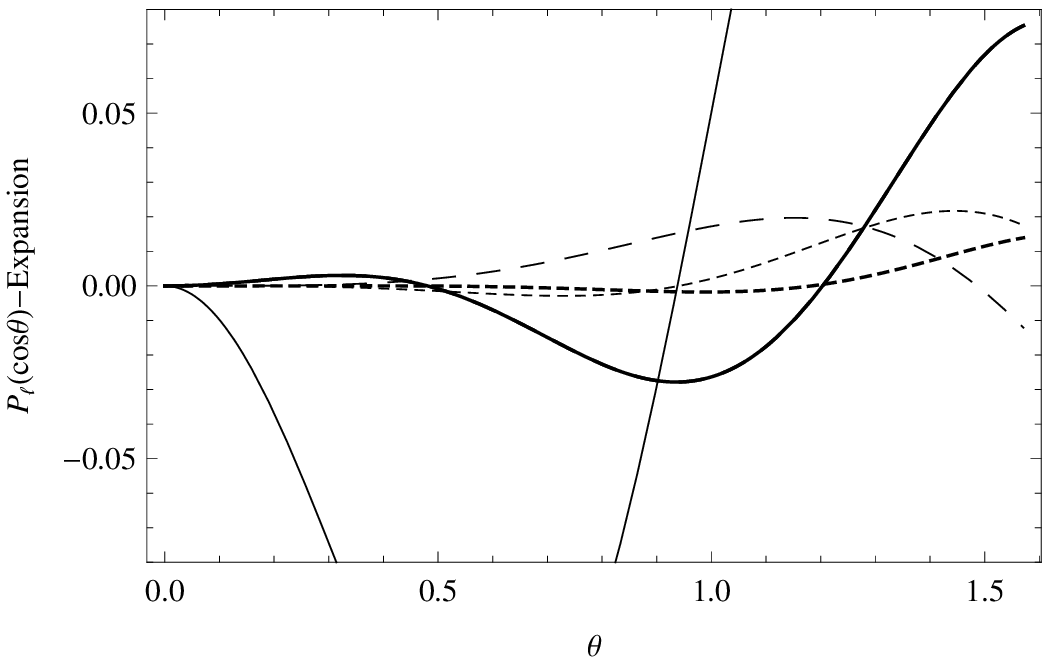}
\includegraphics[width=8.5cm]{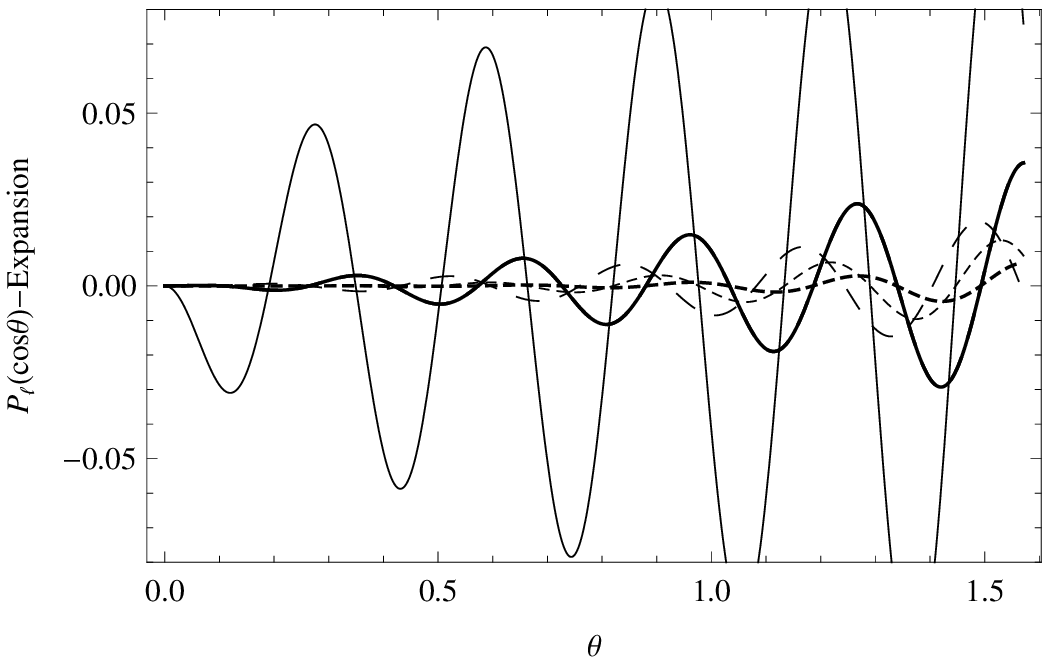}
\caption{Accuracy of the expansions (\ref{Plexpan}) in thin lines and (\ref{Plexpan2}) in thick lines. We plot the difference between $P_{l}(\cos\theta)$ and its approximate forms, at leading order in $\theta$ (i.e. $J_{0}(l\theta)$, solid lines), first order (long dashed lines), second order (short dashed lines) for $\ell=4$ and as a function of $\theta$ (left) and $\ell=20$ (right). } \label{flatsky}
\end{figure*}

This expansion is however not optimal. It was already pointed out in Refs.~\cite{Bond1996,1997ApJ...482....6S,2008PhRvD..78l3506L}
that it can be improved by choosing the argument of the Bessel functions to be ${\etaJPdemi} = L \theta$ with
\begin{equation}
L \equiv \ell+\frac{1}{2}\,,
\end{equation}
instead of $\ell \theta$. The novel expansion can easily be obtained by shifting the argument of the Bessel function in the right hand side of the relation (\ref{Plexpan}). It then reads to second order,
\begin{eqnarray}\label{Plexpan2}
P_{\ell}(\cos\theta)&=&J_{0}({\etaJPdemi})\nonumber\\
&+&\left(\frac{\theta}{2}\right)^2\left[\frac{J_0({\etaJPdemi})}{3}-\frac{J_1({\etaJPdemi})}{6 {\etaJPdemi}}\right] +\dots
\end{eqnarray}
The correction term in $\sim \theta$ of the expansion~(\ref{Plexpan}) has indeed disappeared
and, as a consequence, the first correction to the lowest order of the flat-sky expansion is 
expected to scale as $\ell^{-2}$, (in the sense discussed below).
The accuracy of this mapping is numerically illustrated in Fig. \ref{flatsky} for $\ell=4$ and $\ell=20$,
and shown to be better than the percent level.
As can be appreciated from this expression, from this figure and Fig.~\ref{flatsky2} later on, this change of
expansion point is a very important step that eventually justifies the use
of the flat-sky approximation in the context of precision calculations.
Finally it is to be noticed that 
in order to use correctly the orthogonality relations of the Bessel functions, it is necessary to use the variable
\begin{eqnarray}
Z&\equiv& 2 \tan \left(\theta/2\right)
\end{eqnarray}
instead of $\theta$ since $Z$ runs to infinity when $\theta \to \pi$. The expansion of $P_\ell(\cos \theta)$ 
in function of $\tildeetaJPdemi = L Z $ reads 
\begin{eqnarray}\label{Pexpansionfinale}
\hspace{-0.2cm}&&P_{\ell}(\cos\theta)=J_{0}(\tildeetaJPdemi)\\
\hspace{-0.2cm}&&+\left(\frac{\tildeetaJPdemi}{2L}\right)^2\left[\frac{J_0(\tildeetaJPdemi)}{3} -\frac{J_1(\tildeetaJPdemi)}{6 \tildeetaJPdemi} +\frac{\tildeetaJPdemi J_1(\tildeetaJPdemi)}{3}\right]+\dots\nonumber
\end{eqnarray}

We are then in position to derive the expression of the angular power spectrum in the flat-sky limit.
The starting point is the expression~(\ref{xiexp2}) of the correlation function
where the $\theta$-dependent part of the integrand is expanded as 
$\exp\left[\ii k_{r} (r-r')\right] J_{0}(k_{\perp}r\theta)+{\cal O}(\theta^2)$.
This can then be plugged into Eq.~(\ref{inttheta}), using the expansion~(\ref{Plexpan2})
of the Legendre polynomials and integrated over $\tildeetaJPdemi$, using Eqs.~ (\ref{Ortho1}-\ref{Ortho2}) to give, up
to terms in $\ell^{-2}$,
\begin{widetext}
\begin{eqnarray}\label{ClFBNoIntkperp}
\hspace{-0.4cm}&&C^{(v=0)}_{\ell}=\frac{1}{2\pi}\int \dd r \dd r' \dd k_{r} k_\perp \dd k_\perp \frac{\exp[\ii k_{r}(r-r')]}{r^2}P(k)w(k,r)w^\star(k,r')
\times \\
\hspace{-0.4cm}&&\left[I_{0,0}(L/r,k_{\perp})
-\frac{5}{12r^2}I_{0,1}(L/r,k_{\perp}) 
-\frac{1}{6Lr}I_{1,0}(L/r,k_{\perp}) 
+\frac{L}{12r^3}I_{1,1}(L/r,k_{\perp}) 
+\frac{k_{\perp}}{4r^2}I_{1,1}(k_{\perp},L/r) 
-\frac{\ii k_{r}}{2r}I_{0,1}(k_{\perp},L/r) 
\right],\nonumber
\end{eqnarray}
where the functions $I_{n,p}$ are defined in the appendix \ref{BesselOrtho}.
Integrating on $k_\perp$,  we finally get after integrations by parts 
\begin{eqnarray}\label{ClFB}
C^{(v=0)}_{\ell}&=&\frac{1}{2 \pi}\int\dd r \dd r' \dd k_{r} \frac{\exp[\ii k_{r}(r-r')]}{r^2}
\left[1 +\frac{1}{ 24L^2}\left( D_\perp -6 D_{\perp}^2-4D_\perp^3\right) +\frac{\ii k_r r}{2 L^2}D_\perp^2\right]
P(k)w(k,r)w^\star(k,r')\,,
\end{eqnarray}
\end{widetext}
where $D_\perp\equiv \partial/\partial \ln k_\perp$ acts on all the terms
on its right, and where $k^2=k_{r}^2+L^2/r^2$. This means that we must impose the constraint
\begin{equation} \label{Constraint1}
k_\perp r=L \,.
\end{equation} 

Higher order corrections will introduce higher order derivatives of $P(k)w(k,r)w(k,r')$ with respect to $k_\perp$. 
Since the function on which the derivative acts is only a function of $k$, it is understood in the above expression, and in the rest of this paper, that $\partial_{k_\perp} =(k_\perp/k) \,\partial_k $.
The evolution of the sources is mainly due to the baryon acoustic oscillations and thus $\partial \ln w(k,r)/\partial k \sim \eta_{\lss}$ where $\eta_\lss$ is the mean time of the last scattering surface. 
If the initial power spectrum is a power law, $\partial \ln P(k)/\partial k \propto 1/k$. For super-Hubble scales at the time of recombination, the corrective terms are dominated by the variations of $P(k)$ whereas on sub-Hubble scales, they are dominated by the variation of the transfer function. We thus have
\begin{eqnarray}\label{CondsVariationsk}
\partial_k \ln \left[P(k) w(k,r)w^\star(k,r')\right] \propto
\left\lbrace
\begin{array}{ll}
 1/k  &{\rm if} \,\,\, k\eta_{\lss} \ll 1 \\ 
 \eta_{\lss} &{\rm if}\,\,\, k\eta_{\lss}\gg 1.
\end{array}
\right.
\end{eqnarray}
The scaling of the corrective term can then be inferred easily.
If $\Delta r_\lss$ is the typical width of the last scattering surface, the exponential term in 
the integral on $k_r$ imposes the constraint $k_{r}\Delta r_\lss \lesssim 1$. 
Now, since the power spectrum should be decreasing faster than $k^{-2}$ for large $k$, the
dominant contribution to the integral arises from $k_r \lesssim k_\perp$. We thus conclude that 
$k_r \lesssim {\rm min} (k_\perp,\Delta r_\lss)$ and that we always have $k \approx k_\perp$. 
In the small scale regime, that is for $\ell\gg r_{\lss}/\eta_\lss \approx 50$ 
($r_\lss \equiv \eta_0-\eta_\lss$ being the average comoving distance to the last scattering surface)
then $k \eta_{\lss}\ll 1$. According to Eq.~(\ref{CondsVariationsk}), this means that the first 
corrective term is, in this regime, of order $(\eta_{\lss}/r_{\lss})^2\sim{\cal O}(10^{-4})$. 
In the large scale regime, $\ell\ll r_{\lss}/\Delta r_\lss$, and thus $k\approx k_{r}\approx k_\perp$. 
Since for these modes $\ell \ll r_\lss/ \eta_\lss$, then $k \eta_\lss\ll 1$, and this implies that the 
first corrective term scales approximately as $1/\ell^2$. 

Concerning the higher order expansion of Eq.~(\ref{ClFB}), we see that  
a corrective term of order $1/L^n$ will involve the operator $D_\perp^p/L^n$ where
$p$ can be any integer, which gives formally a systematic way of organizing the
expansion to any order.

To summarize, the precision of the lowest order flat-sky approximation in the context
of CMB calculations is of order $1/\ell^2$ for $\ell \lesssim 50$
and limited to $10^{-4}$ level for $\ell \gtrsim 50$, i.e. they are of
order ${\rm max}(\ell^{-2}, \ell_0^{-2})$ with $\ell_0=50$.  We refer to this first correction as being the correction of order $1/\ell^2$, even if it is limited for small 
scales because of the sub-horizon physics~\footnote{In particular for 
any other observable or physical mechanisms for which the variations  of $w(k,r)$ can be neglected, 
the corrective term would indeed scale as $1/\ell^2$ also on small scales, and this justifies our choice of denomination.}.

We finally note that in some cases the results can actually be further improved by choosing the argument of the Bessel functions to be 
$\etaJPsqrt \equiv \tilde L \theta \equiv \sqrt{l(l+1)}\theta$, since then
\begin{eqnarray}\label{Plexpan3}
\quad P_{\ell}(\cos\theta)&=&J_{0}(\etaJPsqrt)\\
       &&+\left(\frac{\etaJPsqrt}{2 \tilde L}\right)^2\left[\frac{J_0(\etaJPsqrt)}{3} 
       -\frac{2 J_1(\etaJPsqrt)}{3 \etaJPsqrt} \right]+\dots\,.\nonumber
\end{eqnarray}
Indeed, this expansion also removes the corrections scaling as $\ell^{-1}$ in Eq.~(\ref{ClFB}) since it also does not contain terms linear in $\theta$. Now, the flat-sky constraint reads 
\begin{equation}\label{Constraint3}
k_\perp r = \tilde L = \sqrt{\ell(\ell+1)}\,.
\end{equation}
With this choice, it appears that for any transfer function $w(k,r)$ that is constant and sharply peaked, 
i.e. that is such as $w(k,r)=\delta(r-r_\lss)$, and for a scale invariant power spectrum, 
i.e. $P(k)=2 \pi^2 A_s^2 k^{-3}$, the lowest order in the flat-sky expansion takes the form
\begin{eqnarray}\label{ClFBbetter3}
C_{\ell}=\int\frac{\dd k_{r}}{2 \pi} \frac{1}{r_\lss^2} P(k) = \frac{2 \pi A_s^2}{\ell (\ell+1)}\,.
\end{eqnarray}
This is precisely the result that one would have obtained with the exact (or full spherical sky) calculation since
\begin{equation}
\qquad C_\ell = \frac{2}{\pi}\int \dd k k^2 P(k)\left[j_\ell(k r_\lss)\right]^2= \frac{2 \pi A_s^2}{\ell (\ell+1)}\,.
\nonumber
\end{equation}
In other words, with this choice, the leading term of the flat-sky expansion is exact.
For CMB on large scales, this is precisely the case since 
super-Hubble perturbations are frozen, leading to the Sachs-Wolfe plateau.
Given that the initial power spectrum is expected to be almost scale-invariant, we conclude that this  
expansion of $P_\ell(\cos \theta)$ is the best one in the context of CMB computations and
we usually adopt it. We can argue that the choice $k r = L$ is
more compact while the choice $k r = \tilde L$ gives a better formula
only in the thin-shell approximation. In appendix~\ref{AppRelationsFS} we explain how to change the constraint
inside the expressions obtained. In general however, one cannot state which is the best choice and in the following, unless stated otherwise, we
will use only $L$. 

\subsection{Generalized flat-sky expansions}\label{SecRapConv}\label{SecJP}

The derivation of the previous section was limited to the case $v=0$ but, as it appears
clearly from the parameterizations~(\ref{FSparameterizationJP}), it is actually possible to build
a two-parameter family of flat-sky approximations, the relation between which should be examined. 
%


\subsubsection{A one-parameter family of flat-sky approximations}
The aim of this subsection is to explore the consequence of the use of a more general parameterization~(\ref{FSparameterizationJP})
introducing $v$ as a free parameter. Note that we still keep $\psi=\pi$ although it could be reintroduced at this stage too.
Calculations for arbitrary values of $\psi$ are however significantly more complicated and do not lead to any
improved scheme. 
Then, again, the particular cases $v=0$ or $v=1$ correspond to $\hat \vn_z\equiv (0,0,1)$ aligned
either with $\hat \vn$ or $\hat \vn'$ and the case  $v=1/2$ to $\hat \vn_z$
aligned with $(\hat \vn+\hat \vn')/\sqrt{2}$.

We can now follow the same approach as in section~\ref{SecIntuitiveComputation}. 
The expansion in powers of $Z$ (instead of $\theta$) can be performed with the variables  
$\Delta\equiv r-r'$ and 
\begin{equation}
R \equiv r'v + (1-v) r.
\end{equation}
Plugging the expansion~(\ref{Pexpansionfinale}) into Eq.~(\ref{inttheta}) with the definition~(\ref{xiexp2})
of the correlation function, one can perform an expansion in $Z$ and $\Delta$, and 
then, using the orthonormality relations of appendix~\ref{BesselOrtho} we can compute 
the integral on $\tilde {\etaJPdemi}$ in function of the $I_{n,p}(k_\perp,L/R)$.
Eventually the result reads 
\begin{equation}
C^{(v)}_{\ell}=\int\frac{\dd r \dd r' \dd k_{r}}{2 \pi} \frac{\exp(\ii k_{r} \Delta)}{R^2} \mO^{(v)}_{k_{\perp}}P(k)w(k,r)w^\star(k,r')\,,
\end{equation}
where $ \mO^{(v)}_{k_{\perp}}$ is an operator that applies on the right part of the previous expression with

\begin{eqnarray}
\hspace{-1.0cm}&& \mO^{(v)}_{k_{\perp}}=1 +\frac{ D_\perp}{24L^2}-[(4-3 f_{1/2})R+f_{1/2}f_{0}\Delta]\frac{D_\perp^3}{ 24RL^2}\nonumber\\
\hspace{-1.0cm}&&-[f_{1/2} f_{0}\Delta+3 R f_{0}^2]\frac{ D_\perp^2}{
   12RL^2}+\ii k_r(4 f_{0} R +f_{1/2}\Delta)\frac{D_\perp^2}{8L^2}
\end{eqnarray}
where $f_{1/2}\equiv 4v(1-v)$, $f_{0}=1-2v$ and
\begin{equation}
D_\perp\equiv k_\perp\frac{\partial}{\partial
  k_\perp}=\frac{k_\perp^2}{k}\frac{\partial}{\partial k}=\frac{L^2}{kR^2}\frac{\partial}{\partial k}\,.
\end{equation}
Besides the $v=0$ case for which this expression simplifies, this is also the case for the symmetric choice, $v=1/2$,
for which it leads to,
\begin{widetext}
\begin{equation}\label{ClJP}
C^{(v=1/2)}_{\ell}=\int\frac{\dd r \dd r' \dd k_{r}}{2 \pi} \frac{\exp(\ii k_{r} \Delta)}{R^2}\left[1 +\frac{1}{ 24L^2}\left( D_\perp -D_\perp^3\right) +\frac{\ii k_r \Delta }{8  L^2}D_\perp^2\right] P(k)w(k,r)w^\star(k,r')\,.
\end{equation}
\end{widetext}
The lowest order of this expression matches the one derived in Ref.~\cite{Bond1996}.
We remind that the constraint $k_\perp R = L$ must be satisfied and that this expression is valid only 
up to order $L^{-2}$. In the general case for which $v\not=1/2$, at lowest order in the flat-sky expansion, the expression would remain formally the same, considering that $R$ would then be given by $R=r' v + (1-v)r$. We will compare these different parameterizations in 
the following paragraph. However, if we also take into account the corrections, the choice of $v$
would change the expression of the flat-sky expansion. In appendix~\ref{AppRelationsFS} we detail 
how the different corrections obtained are consistent with one another.

%

\subsubsection{Breaking of statistical isotropy and off diagonal contributions}

The existence of mathematically equivalent flat-sky approximations may appear surprising at first view
(at least it surprised the authors) but it can be fully understood when one addresses the construction
of the correlators in harmonic space \textsl{not assuming the statistical isotropy of the sky.}

Indeed an important consequence of the flat-sky approximation is, because
a particular direction has been singled out, to break
the statistical isotropy of the sky. There is therefore no reason to have,
$C_{\ell_1 m_1 \ell_2 m_2}=C_{\ell_1}\delta_{\ell_1\ell_2}\delta_{ m_1m_2}$
at any order in flat-sky approximation where
$C_{\ell_1 m_1 \ell_2 m_2}$ is the ensemble average of the product of
two spherical harmonics coefficients. 

\begin{figure}[!htb]
 \includegraphics[width=8.5cm]{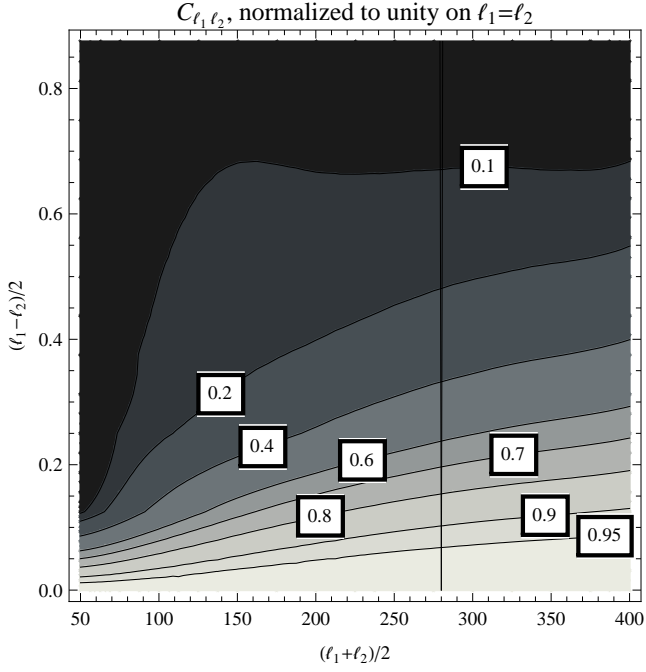}
 \caption{$C_{\ell_1 \ell_2}/C_{\ell_+ \ell_+}$ with $\ell_+=(L_1+L_2)/2-1$. Most of the signal 
 is localized on the diagonal $L_1=L_2$. The choice of the path of integration
 in this space leads to different flat-sky expansions, that is to different choices of
 the flat-sky direction $\hn$ or equivalently of the parameter $v$.
 For $v=0$ or $v=1$, the lowest order of Eq.~(\ref{ClFB}) can be recovered by integrating 
 on an horizontal or a vertical line in the $(L_1,L_2)$ plane. As for $v=1/2$, the lowest order of the expression (\ref{ClJP})  is recovered by integrating on the line $L_1+L_2= $const. The vertical black line(s) is the superposition of those various integration paths for $L=280$. On the plot they are hardly distinguishable.
}\label{flatsky3}
\end{figure}

In terms of the two-point angular correlation functions we have in general,
\begin{equation}\label{Cl1l2m1m2Def}
C_{\ell_1 m_1 \ell_2 m_2}\equiv \int \dd \hat \vn_1 \dd \hat \vn_2 \xi(\hat \vn_1,\hat \vn_2)Y_{\ell_1}^{m_1}(\hat \vn_1) Y_{\ell_2}^{\star m_2}(\hat \vn_2)\,,
\end{equation}
where $\xi(\hat \vn_1,\hat \vn_2)$ is the correlation function of two given directions. The difference with
Eq.~(\ref{defxi}) is that we have not assumed isotropy. 
If we choose these two directions to be close to a common
direction $\hat \vn$, then we can expand the spherical harmonics of Eq.~(\ref{Cl1l2m1m2Def}) around that
direction which is chosen to be aligned with the azimuthal direction and this step breaks explicitly the isotropy of the problem. 
At lowest order, this gives (see appendix~\ref{AppYlmAsymptotics})
\begin{equation}
Y_\ell^m(\hat \vn)\simeq \sqrt{\frac{2 \ell+1}{4 \pi}}(-1)^m J_m(\ell \theta) e^{\ii m \varphi}.
\nonumber
\end{equation}
It implies that \begin{equation}
C_{\ell_1 m_1 \ell_2 m_2}\simeq \delta_{m_1 m_2}C_{\ell_1 \ell_2},
\end{equation}
as a consequence of the preservation of the statistical rotational invariance around the particular direction $\hn$.
The term $C_{\ell_1 \ell_2}$ is explicitely given by
\begin{eqnarray}\label{Cl1l12expr1}
&&C_{\ell_1 \ell_2}\equiv\frac{1}{2 \pi}\int \dd r_1 \dd r_2 \dd k_\perp k_\perp \dd k_r P(k) w(k,r_1)w^\star(k_,r_2)\nonumber\\
&&\quad\times\exp[\ii k_r(r_1-r_2)]\frac{\delta(k_\perp r_1-L_1)\delta(k_\perp r_2-L_2)}{\sqrt{L_1 L_2}}
\end{eqnarray}
where we remind that $L\equiv \ell+1/2$ (or  $= \sqrt{\ell(\ell+1)}$).
Performing the integrals on $r_1$ and $r_2$ leads to
\begin{eqnarray}\label{Cl1l12expr2}
C_{\ell_1 \ell_2}&\equiv& \frac{1}{2 \pi}\int \frac{\dd k_\perp}{ k_\perp} \dd k_r P(k) \exp\left[\ii (L_1-L_2)\frac{k_r}{k_\perp}\right]\nonumber\\
&&\qquad \times\frac{w\left(k,\frac{L_1}{k_\perp}\right)w^\star\left(k,\frac{L_2}{k_\perp}\right)}{\sqrt{L_1L_2}}\,. 
\end{eqnarray}
In Fig~\ref{flatsky3} we present $C_{\ell_1 \ell_2}$ in the space of $(\ell_1+\ell_2)/2$ and $(\ell_1-\ell_2)/2$.
The main power is carried by multipoles such that
$\ell_1\approx\ell_2$. The difference between the different possible flat-sky approximations
lies precisely in the existence of (small) off-diagonal terms.

\subsubsection{Recovering the different flat-sky approximations}

Interestingly, each $C_\ell^{(v)}$  can be recovered by a proper integration of
$C_{\ell_1 \ell_2}$, given by Eq.~(\ref{Cl1l12expr1}), in the $(\ell_1,\ell_2)$-plane.
Each path of integration corresponds to a way to relate
the correlation function $\xi(\hat \vn_1,\hat \vn_2)$ to $\xi(\theta)$ defined in Eq.~(\ref{defxi}). 

For instance, when $v=0$, the expression~(\ref{ClFB}), which is then the lowest order part
of the expression~(\ref{ClJP}), can be recovered  by integrating along the path
$L_1$=const., i.e. as
\begin{equation}
L_1 C_{\ell_1} = \int \dd L_2 \sqrt{L_1 L_2} C_{\ell_1 \ell_2}.
\end{equation}
The integration along $L_2$=const. corresponds to the case $v=1$
and gives by symmetry the same result, which was expected since it 
corresponds to exchanging $\hn_1$ and $\hn_2$.

The symmetric case $v=1/2$ can be recovered by
integrating on the path $L_1+L_2$=const. Making the change of variables
\begin{equation}\label{Changel1l2}
L_{+}=\frac{L_1+L_2}{2}\,,\qquad L_{-}=\frac{L_1-L_2}{2}\,,
\end{equation}
we thus obtain
\begin{eqnarray}
C_{\ell_1} &=& \int \frac{\dd L_-}{2} \frac{\sqrt{L_1 L_2}}{L_+} C_{\ell_1 \ell_2} \nonumber\\
&=& \int \frac{\dd L_-}{2} \sqrt{\left(1-\frac{L_-^2}{L_+^2}\right)} C_{\ell_1 \ell_2} \,.
\end{eqnarray}

Actually, for a general $v$, the lowest order of the expression~(\ref{ClJP}) is
recovered by integrating on the path of equation 
\begin{equation}
L_2/L_1=(v-1)/v
\end{equation}
in this plane. The case of $\psi\ne \pi$ in Eq.~(\ref{FSparameterizationJP}) leads to similar constructions but with 
slightly more complicated integration paths. To show it let us reintroduce $\psi$ in Eq.~(\ref{FSparameterizationJP}). Then the angular separation
$\tilde\theta$ of the directions $\hn$ and $\hn'$ reads
\begin{equation}
\tilde\theta^2=(v^2+(1-v)^{2}-2v (1-v)\cos\psi)\theta^2
\end{equation}
and the effective radius $R$ (as it appears in the argument of $J_{0}$) is 
\begin{equation}
R^2=(r^2v^2+r'^2(1-v)^{2}-2r r' v (1-v)\cos\psi)\theta^2
\end{equation}
with the relation
\begin{equation}
L\tilde\theta=k_{\perp}R.
\end{equation}
The relation between $L$, $L_{1}$ and $L_{2}$ then follows from the relations $L_{1}=r k_{\perp}$ and
$L_{2}=r' k_{\perp}$. It reads,
\begin{equation}
L^2=\frac{v^2 L_{1}^2+(1-v)^2 L_{2}^2-2v (1-v)\cos\psi L_{1}L_{2}}{v^2+(1-v)^{2}-2v (1-v)\cos\psi} \end{equation}
which describes the arc of an ellipse in the $(L_1,L_2)$ plane. All these approximations are a priori
of similar precision. Note however that when $\psi$ is small, $\theta$ should be large in order to keep
$\tilde\theta$ fixed making the various expansions less precise. It then corresponds to a very squeezed ellipse
in the $(L_1,L_2)$ plane.

\section{Two limit cases}\label{SecApproximations}


It is worth remarking that in either Eq.~(\ref{ClFB}) or more
generally in Eq.~(\ref{ClJP}), the computation of power spectra at
leading order involves a genuine 3D numerical integration. It is therefore numerically 
less favorable than exact calculations (which involves only two integrals~\cite{1996ApJ...469..437S,1997PhRvD..56..596H} ~\footnote{More precisely, the exact result has
three integrals as well, but the integral on $r$ is squared and numerically we have effectively only two integrals.}.)
It is just more transparent since it does not require the computation of spherical Bessel functions, and geometrically more transparent 
since the results are presented in a Cartesian form. 

Depending on the physical situation, it is however possible to introduce simplifications that will
make the computations faster.

\subsection{The Limber approximation}\label{limbersec}

Let us first consider the situation in which the sources stretch in a wide range
of distances $\Delta r$ and vary smoothly. As long as $\ell/r \times \Delta r \gg 1$, it implies
that $k_\perp \Delta r \gg 1$. As a consequence, the contributions in the integral 
on $k_r$ are approximately ranging from $-1/\Delta r$ to $1/\Delta r$, that is in a range 
of values where $k=\sqrt{k_\perp^2 + k_r^2}\simeq  k_\perp$ and all functions except 
the exponential in the integrand can be considered constant.

The Limber approximation then consists in replacing
\begin{equation}
\int \dd k_r \exp[\ii k_r (r-r')]\nonumber
\end{equation}
by $ 2 \pi \delta {(r-r')}$ so that the integral on $r'$ can then be performed
trivially. We finally obtain, 
at lowest order in powers of $\ell^{-1}$
\begin{equation}\label{ClLimber}
C_{\ell}=\int \dd r \left|\frac{w(k,r)}{r}\right|^2P(k)\,,
\end{equation}
with the Limber constraint $k r = L$. The first corrections
scale as $L^{-2}$ and have been computed in Ref.~\cite{2008PhRvD..78l3506L}. 
We can recover this result from Eq.~(\ref{ClJP}), by expanding all functions of $k$ around $k_\perp$ as
\begin{equation}
f(k)=f(k_\perp)+\frac{k_r^2}{2 k_\perp}\left.\frac{\partial f(k)}{\partial k}\right|_{k=k_\perp}+\dots
\nonumber
\end{equation}
This will result in a term proportional to $k_r^2$ which can be handled using
\begin{equation}\label{kr2inLimber}
\int \dd k_r (ik_r)^n\exp[\ii k_r \Delta]= 2\pi \delta^{\{n\}} (\Delta)\,,
\end{equation}
where $\delta^{\{n\}}$ is the $n$-th derivative of the Dirac distribution.
The term proportional to $k_r$ in Eq.~(\ref{ClJP}) can be handled in the same way
and gives $\delta'(\Delta)$. Integrating by parts in $\Delta$  removes the derivatives of 
the Dirac distributions, using that $\partial R/\partial r= 1-v$ and $\partial R/\partial r'=v$. 
Then, the integral on $r'$ is trivial, because of the Dirac distributions, and one can then perform
an integration by parts in $r$ in order to reshape the result. We finally obtain
\begin{widetext}
\begin{equation}\label{ClLimber2}
\hspace{-0.2cm}C^{(v=1/2)}_{\ell}=\int  \frac{\dd r}{r^2}\left\{P(k)\left|w(k,r)\right|^2+ \frac{D}{2 L^2} \left[r\left|\frac{\partial \left[w(k,r)/\sqrt{r}\right]}{\partial \ln r}\right|^2P(k)\right]-\frac{D^2}{2 L^2} [w(k,r)^2P(k)]-\frac{D^3}{6 L^2} [w(k,r)^2P(k)] \right\}
\end{equation}
\end{widetext}
where $D^n\equiv k^n \partial /\partial k^n$ and we recover the result
of Ref.~\cite{2008PhRvD..78l3506L}. The Limber approximation is not well suited for CMB computation since its 
hypothesis is not satisfied during recombination. However, when it comes to the 
contributions of the reionization era, sources stretch in a wide range of distances 
and the Limber approximation can be used. Special care must be taken though because 
the source have a directional dependence and are not simple scalar valued sources. 
The main sources are ({\it i}) the Sachs-Wolfe contribution
$\delta_r/4 + \Phi$ where $\delta_r$ is the 
density contrast of the radiation ({\it ii})  the late variation $\Phi'$ of
the gravitational potential, and ({\it iii})  the Doppler 
contribution $n^i \partial _ i v$, where $v$ is the scalar part of the baryon velocity. 
A naive estimation would lead to think that the Doppler effect is the most
significant contribution since $v \sim (r_\lss-r)$ while the intrinsic Sachs-Wolfe term remains of order one. However, in Fourier space this leads to a term 
proportional to $k_r v$, and it cannot be dealt with the naive replacement $k_r \to 0$ since this would vanish. 
Instead, it should be treated using Eq.~(\ref{kr2inLimber}) and it leads to replace the Doppler source by $\ii \dd v/\dd r$ (if we ignore the derivatives of $1/r$) before taking the lowest order of the Limber approximation, and this then gives
a contribution of order unity as well. Physically, this means that the modes which favor the Doppler 
effect are those aligned with the line-of-sight $n^i$, but the contribution of these modes is 
further suppressed in the Limber approximation for sources stretching in a wide range of distances. 
See Section~\ref{SecSpectrumGeneral} below for a discussion on the flat-sky expansion 
in general in the case of sources with intrinsic directional dependence.  
Furthermore, if we decide to consider the late Integrated Sachs-Wolfe (ISW) effect which is due to the 
variation of the gravitational potentials, the Limber approximation is satisfactory as well. 
We shall not consider any of these in this article and focus our attention to the effects 
that occur during the recombination.

\subsection{The thin-shell approximation}\label{thinsec}

A useful approximation can be derived when the sources are contributing 
only in a thin range of distances.

Indeed, in such a case the factors $1/r^2$ or $1/R^2$ can be replaced by $1/r_\lss^2$, where $r_\lss$ is the
mean distance of the sources each time one has to compute $k$ (e.g. $k=\sqrt{k_{r}^2+L^2/R^2}$ is replaced by
$k_\lss\equiv\sqrt{k_{r}^2+L^2/r_\lss^2}$). The integrals on $r$ and $r'$ in the lowest order term of expression~(\ref{ClJP}) can be factorized and we obtain the thin-shell flat-sky approximation
\begin{equation}\label{ClThinShell}
\hspace{-0.0cm}C_{\ell}=\frac{1}{r_\lss^2}\int\frac{\dd k_{r}}{2 \pi} \left|\int \dd r \exp(\ii k_{r}r)w(k_\lss,r)\right|^2P(k_\lss)\,.
\end{equation}
In the full-sky computation, this factorization is automatic since there is also
an integral over $r$, which is then squared, and another integral over $k$.  
In general, the flat-sky expansion breaks this property, and it is recovered in 
the thin-shell approximation. 

Estimating the error introduced by the thin-shell approximation is actually difficult and quite model dependent. First the error introduced by replacing the factor $1/R^2$ by
$1/r_\lss^2$ will be very limited if $r_\lss$ is taken in the middle
of the LSS and should be much less than $\Delta r_\lss/r_\lss$. However we also need to estimate the error introduced by
the approximation of the flat-sky constraint. Depending on the $k$ dependence of $S(k,r)\equiv w(k,r)w(k,r')^\star
P(k)$, this will lead to two extreme cases. If it is a pure power law
and depends only on $k$, then $S(k)\simeq S(k_\lss)- (R-r_\lss)/r_\lss D_\perp S(k_\lss)$
and again if $r_\lss$ is chosen in the middle of the LSS, after
integration on $r$ and $r'$, the error would be much smaller than $\Delta r_\lss/r_\lss$. However if the
sources are purely oscillatory with frequency $k$, which is the case
for the small scales, that is if $S(k,r)
\propto \cos(k r)$ for instance, the relative error introduced would be of
order $\Delta r_\lss/r_\lss$. So for small scales, this is much larger than the first
corrections in the expression~(\ref{ClFB}) which scale as ${\rm max}(\ell^{-2}, \ell_0^{-2})$ with $\ell_0\equiv
r_\lss / \eta_\lss\simeq 50$. We shall see in
the next section that it is in principle comparable on small scales to the larger corrections
introduced when we consider the non-scalar nature of the sources which
are of order ${\rm max}(\ell^{-1}, \ell_0^{-1})$. However, in practice the error introduced in the
thin-shell approximation is smaller since $\Delta r_\lss<
\eta_\lss$ and also because the sources are not purely oscillatory and
converge to a power-law on small scales thanks to viscous effects, thus reducing further the error made.

We can also try to compute  corrective terms to the thin-shell
approximation. This correction is obtained by
expanding $R$ around $r_\lss$ but also $k$ around $k_\lss$ and is given
at lowest order of this expansion by 
\begin{eqnarray}\label{CorrectThinshell}
\hspace{-0.35cm}&&\delta C_{\ell}^{\rm thin\,\,shell}=\\
\hspace{-0.35cm}&&\int\frac{\dd r \dd r' \dd k_{r}}{2 \pi}
\frac{\exp(\ii k_{r} \Delta)}{r_\lss^2} \delta \mO P(k_\lss)w(k_\lss,r)w^\star(k_\lss,r')\,,\nonumber
\end{eqnarray}
with 
\begin{equation}
\delta \mO \equiv \left[1-\frac{(R-r_\lss)}{r_\lss}\left(2+D_\perp\right)\right].
\end{equation}
Since $R-r_\lss =v[r'-r_\lss] +(1-v)[r-r_\lss]$ the integrals on $r$ and $r'$ in the expression of this correction can
also be expressed as a sum of factorized integrals, which means that
numerically it corresponds effectively to sums of two-dimensional integrals.
On large scales the operator $D_\perp$ acts mainly on $P(k_\lss)$
since it contains the dominant $k$ dependence (see the discussion in section~\ref{SecIntuitiveComputation}). However on small scales
the dependence in $k$ is dominated by the source terms $w(k_\lss,r)$ and
$w^\star(k_\lss,r')$, and this lowest order correction is not valid
given the numerous oscillations of the sources. However for a source
which depends purely on $k r$, as is roughly the case for the baryon
acoustic oscillations, then on small scales it depends
nearly on $k_\perp r$ since $k\simeq k_\perp$, and the error
introduced by replacing $k$ with $k_\lss$ can be seen as an error in the placement
of the distance $r$ at which the source is emitting. In that
limit case, everything happens as if the visibility function
contained in the expression of the source $w(k,r)$ was slighlty
distorted when performing the thin-shell approximation. In practice, we shall not correct for this since the source
is not purely depending on $k r$ on small scales. This means that
we should make the operator $D_\perp$ contained in $\delta \mO$ act
only on $P(k_\lss)$ for our practical purposes when using Eq.~(\ref{CorrectThinshell}) to correct
for the thin-shell approximation that we take.

 In the context of CMB, the use of the thin-shell approximation requires 
to rewrite the source in order to localize the physical effects on the LSS. In practice,
the terms involving the gravitational potential $\Phi$, i.e. of the type 
$n^i \partial_i \Phi$~\cite{1997PhRvD..56..596H,1995ApJ...455....7M}, whose contribution would stretch from the LSS up to now, are replaced by $\dd \Phi/ \dd r-\partial \Phi/\partial r$. This clearly splits the effect 
into an effect on the LSS ($\dd \Phi/ \dd r$), the Einstein effect, and an integrated effect 
($\partial \Phi/\partial r$) which is negligible for a matter dominated universe since the potential is then constant. 
Should we consider the effect of the cosmological constant on the variation of the gravitational potential, 
then we could use the Limber approximation discussed in the previous paragraph, 
and sum the resulting $C_\ell$ to the contribution of the LSS.

\begin{figure*}[!htb]
\includegraphics[width=8.5cm]{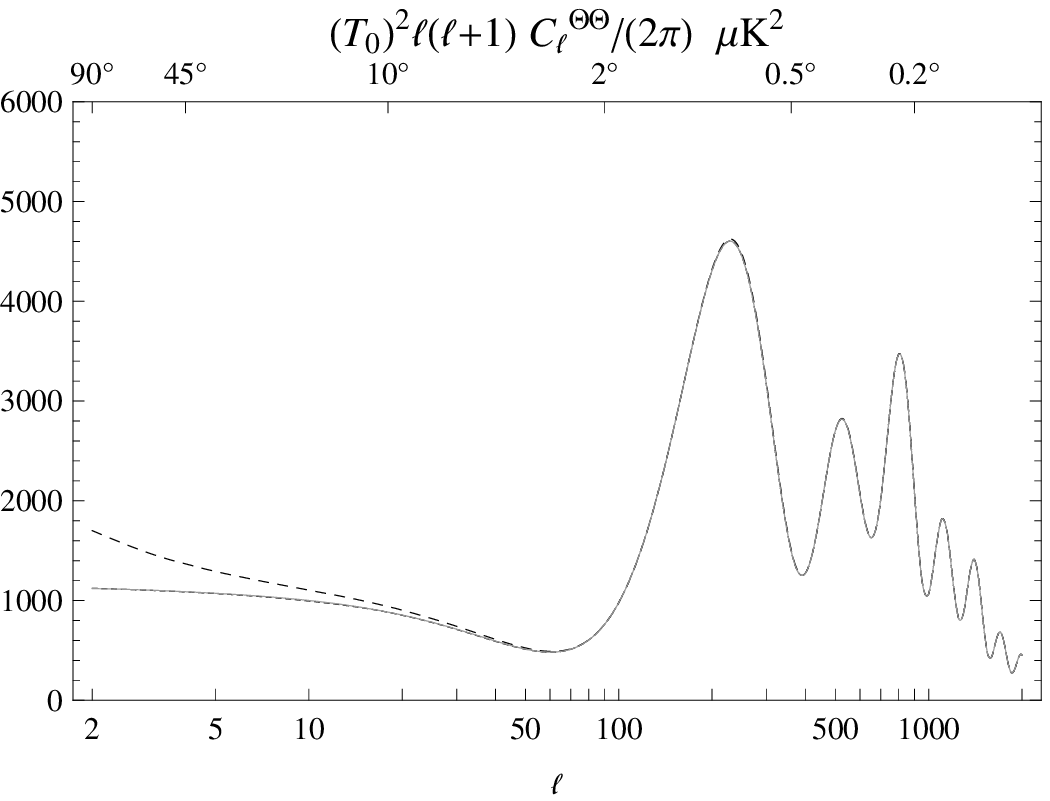}
\includegraphics[width=8.5cm]{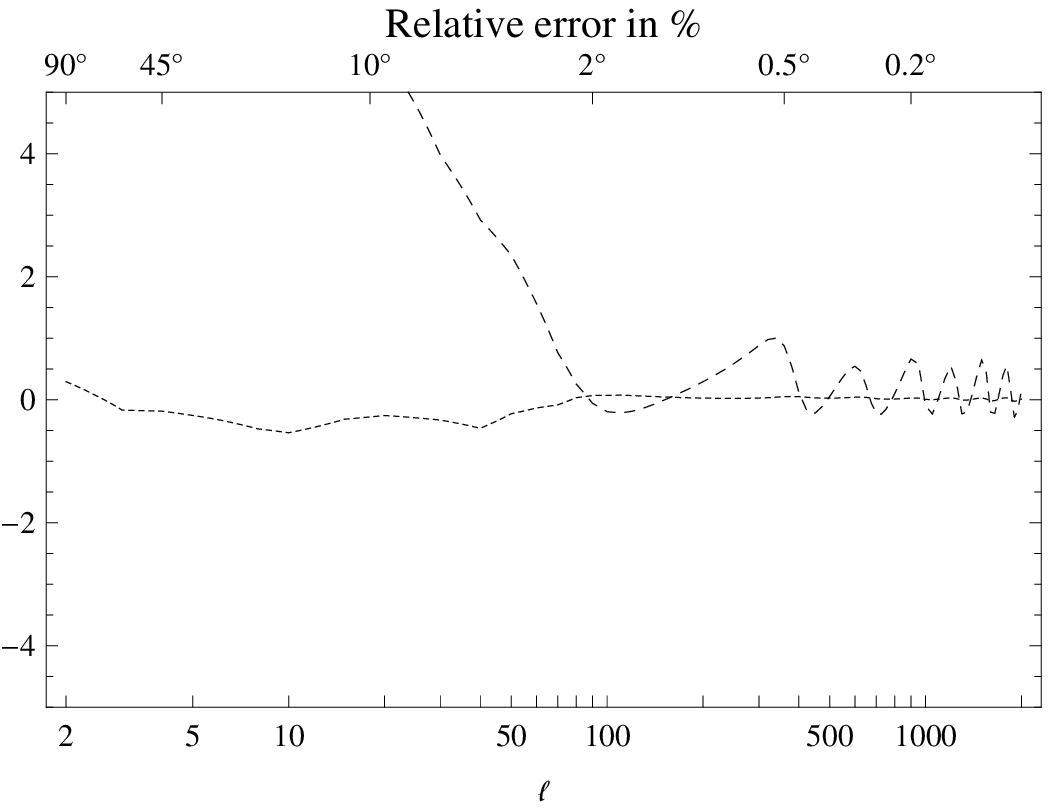}
\caption{{\em Left:} Comparison of the flat-sky approximations
(with $k_\perp r=\ell$ in dashed line and with $k_\perp r = \sqrt{\ell(\ell+1)}$
in dotted line) to the  exact computation (solid line). We consider the standard cosmology
with an instantaneous recombination and ignore all effects but the intrinsic 
Sachs-Wolfe effect ($\Theta=\delta_{{\rm r}}/4 + \Phi$). 
{\em Right:} The relative errors of these two flat-sky approximations with respect to the exact computation.
The first method is limited to a 1\% relative error above $\ell=100$, 
as discussed in Section~\ref{SecIntuitiveComputation}, whereas the second one is much 
better since the first corrections scale as $\ell^{-2}$.
}\label{flatsky2}
\end{figure*}

In Fig.~\ref{flatsky2}, we present the flat-sky approximation for an instantaneous recombination, including 
only the intrinsic Sachs-Wolfe effect in order for the source to be purely scalar. 
The corrections in the thin-shell approximation can be read from the expression~(\ref{ClJP}). 
However for practical purposes, the derivatives with respect to $k_\perp$ need to be converted 
into derivatives with respect to $k$ so that their action on $w(k,r)w(k,r')^\star P(k)$ is clearer. 
However, as we shall see in Section~\ref{SecSpectrumGeneral}, for realistic purposes, 
the sources are not a pure scalar and have an intrinsic geometric dependence so that
the first correction will actually scale as $L^{-1}$.

It is interesting to remark that all flat-sky expansions will lead to the same expressions at lowest
order in the thin-shell approximation and in the Limber approximation, since then, as we have already seen, the lowest order does not depend on
the choice of $v$ (nor $\psi$). In the case of the thin-shell this can be understood from
Fig.~\ref{flatsky3}. The thiner the shell, the more the function $C_{\ell_1\ell_2}$ is peaked
on the diagonal $L_1=L_2$ and the less the integral depends on the line of integration in the $(L_1,L_2)$
plane. On Fig.~\ref{flatsky3}, we plot two contours of integration corresponding
respectively to $v=0$ and $v=1$, and it can be understood that the integrals obtained
on these contours cannot be very different. The Limber case is similar since  the
factor
\begin{equation}
\int \dd k_r \exp\left[\ii (L_1-L_2)\frac{k_r}{k_\perp}\right]
\end{equation}
in Eq.~(\ref{Cl1l12expr2}) is approximated to be $2 \pi k_\perp\delta(L_1-L_2)$.


\section{Effect of non-scalar source terms} \label{SecSpectrumGeneral}

So far, we have assumed that the transer function was scalar, in the sense that
it was a function $w(k,r)$ that does not depend on $\hk$ and $\vn$, that is
the expansion~(\ref{Expsources}) contained only a term $w_{00}$.
In general, this is not the case since scalar perturbations
involve $m=0$ terms with $\ell=1$ (Doppler effect) while vector
perturbations and gravity waves generates terms with $m=1$ and
$m=2$ respectively.

In order to take this dependence into account in the flat-sky analysis, one needs 
to compute Eq.~(\ref{xiexp2}) with the source~(\ref{Expsources}). 
From the parameterization (\ref{FSparameterizationJP}) with the choice $\varphi=0$, we deduce that
\begin{equation}\label{Devkn}
\hat \vk.\hat \vn= \frac{1}{k}\left[k_r + k_\perp \frac{\theta}{2} \cos(\beta)+{\cal O}(\theta^2)\right]\,
\end{equation}
on small scales, where we remind that $\beta$ is defined in the
parameterization~(\ref{paramk}) of $\vk$.

\subsection{Lowest order flat-sky expansion}

As long as we consider only scalar perturbations, the source term
will contain $w_{\jj m}$ terms with $m=0$. Note that this is
different from our previous assumption that only $w_{00}$
was not vanishing.

The expansion~(\ref{Expsources}) contains only terms in 
 $Y_{\vk}^{\jj 0}(\hat \vn)$ which are proportional to $P_\jj(\hat \vk .
 \hat \vn)$. At lowest order in $\theta$, $P_{\jj}^m(\hat \vk.\hat \vn)$ depends only on $k_r/k$
so that the flat-sky expansion remains unchanged as long as we make
the replacement
\begin{equation}
w(k,r)\to \sum_{\jj} \ii^\jj w_{\jj 0}(k,r)P_\jj(k_r/k)\,.
\end{equation}

In conclusion, the formal expression of the flat-sky approximation at lowest
order remains unchanged for cosmological scalar perturbations (which are the dominant
sources of CMB anisotropies).

\subsection{First correction}

\begin{figure*}[!htb]
\includegraphics[width=8.5cm]{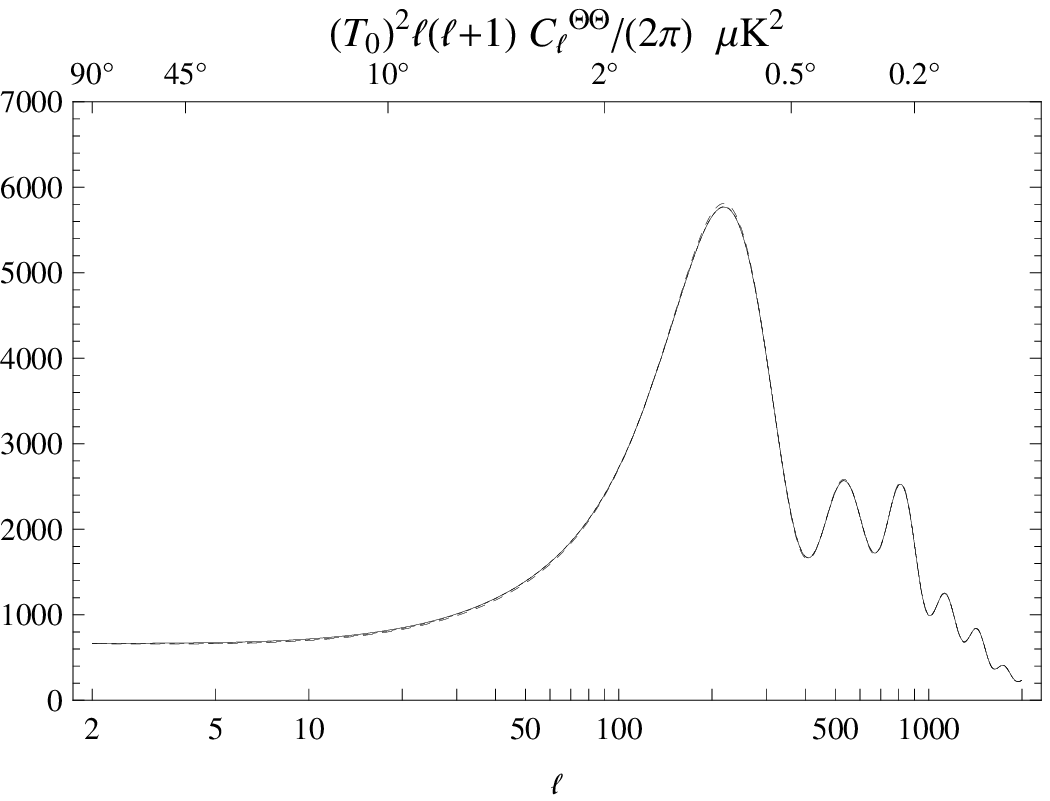}
\includegraphics[width=8.5cm]{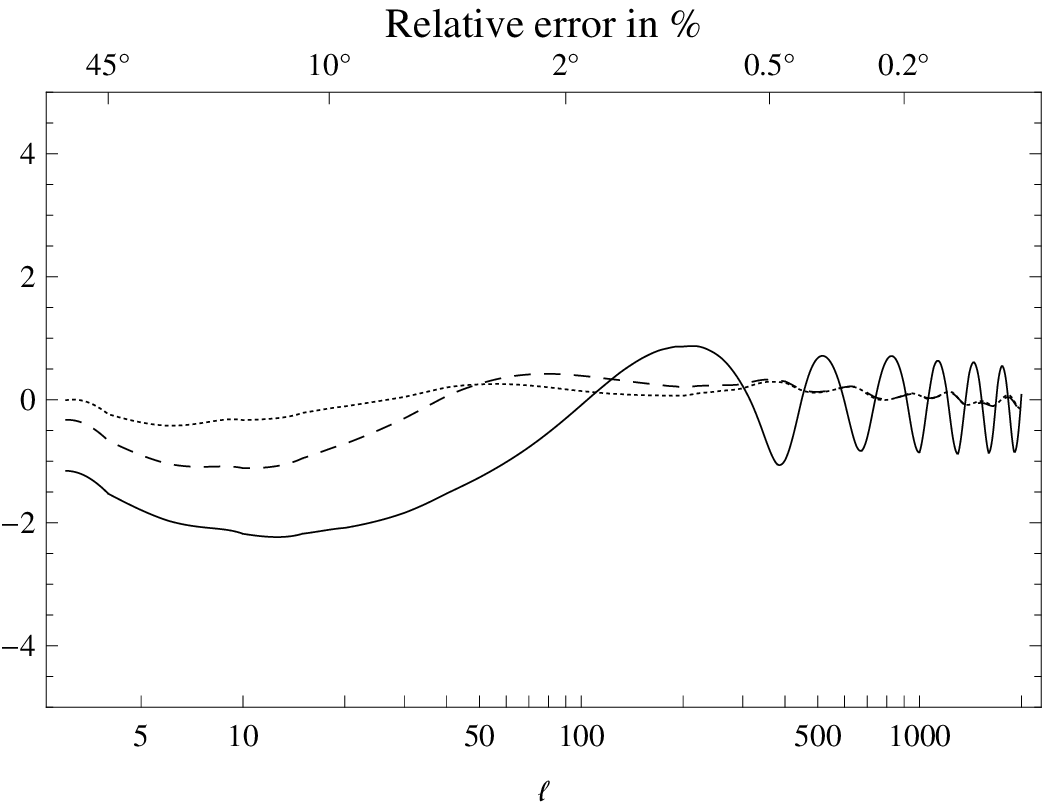}
\caption{{\em Left:} Comparison of the flat-sky approximations with the constraint $k_\perp r = \sqrt{\ell(\ell+1)}$
while including only the lowest order in the expression~(\ref{ClJP2}) (dashed
line) or adding its correction of order $\ell^{-1}$ (dotted line) to the exact
computation (solid line,  standard cosmology, ignoring only the late time ISW effect). 
The calculations in the flat-sky approximations have been done using the thin-shell
approximation. 
\\
{\em Right:} The relative errors of these two approximations with respect
to the exact computation in respectively continuous for the lowest
order expression and dashed line when including the first correction in $\ell^{-1}$. In dotted
line we also compute the error when adding on top of the order
$\ell^{-1}$ correction, the correction of Eq.~(\ref{CorrectThinshell})
due to the thin-shell approximation. The lowest order is limited to a 1\%
relative error beyond $\ell=100$ whereas the first correction
increases substantially the presicion on small scales and the
correction for the thin-shell approximation improves also the largest scales.}\label{CorrectiveTermDoppler}
\end{figure*}

We have seen that for scalar field sources, the first correction scale as $\ell^{-2}$. However, for the general case, 
the first correction arises from  the first correction in (\ref{Devkn}) and scales as $\ell^{-1}$. Since the dominant 
effects come from $\jj=0$ and $\jj=1$, we can ignore the contribution coming from $\jj=2$ in the
computation and there is no contribution for $\jj \ge 3$.
Thus, we restrict to 
\begin{eqnarray}
w(\vk,\hat \vn ,r)&=&[w_{0 0}(k,r)+\ii \frac{k_r}{k}w_{1 0}(k,r)]\nonumber\\
&&+\frac{\theta}{2} \left[\ii \frac{k_\perp}{k} w_{1 0}(k,r) \cos \beta \right]+{\cal O}(\theta^2)\,.
\end{eqnarray}
In the two-point correlation function, the first correction will arise from the product of the first order 
of $w(\vk,\hat \vn ,r)$ with its lowest order term. Using the symmetry of the integral 
in $k_r$ the first corrective term due to the geometry of the sources is thus
\begin{eqnarray}
&&\hspace{-0.5cm}w(\vk,\hat \vn ,r)w^\star(\vk,\hat \vn' ,r')\simeq\\
&&\hspace{-0.5cm}\left[w_{0 0}(k,r)+\ii \frac{k_r}{k}w_{1 0}(k,r)\right]\left[w_{0 0}(k,r')+\ii \frac{k_r}{k}w_{1 0}(k,r')\right]^\star\nonumber\\
&&\hspace{-0.5cm}+\frac{\theta}{2}\ii \cos \beta \frac{k_\perp}{k} \left[  w_{0 0}(k, r) w^\star_{1 0}(k,r') + w_{1 0}(k, r) w_{0 0}^\star(k,r') \right]\nonumber\\
&&\hspace{-0.5cm}\equiv \overline{w w^\star}(\vk,r,r')+\theta \ii \cos\beta \left( w w^\star \right)^{(1)}(\vk,r,r')\,.\nonumber
\end{eqnarray}
In order to go from Eq.~(\ref{Theta}) to Eq.~(\ref{xiexp2}), the integral over $\beta $ will give a
term $-J_1[k_\perp R Z]$ instead of the previous term $J_0[k_\perp R Z]$ 
(due to the factor $\ii \cos \beta$). Following the exactly same method as in Section~\ref{SecJP},
we obtain the flat-sky expression  with the first correction included
\begin{eqnarray}\label{ClJP2}
C_{\ell}&=&\int\frac{\dd r \dd r' \dd k_{r}}{2 \pi} \frac{\exp[\ii
  k_{r}(r-r')]}{R^2}\Big\{P(k)\overline{w w^\star}(\vk,r,r')\nonumber\\
&&\qquad\quad-\frac{D_\perp}{k_\perp^2 r}  \left[k_\perp P(k)\left(w w^\star\right)^{(1)}(\vk,r,r')\right]\Big\},
\end{eqnarray} 
where we can take either the flat-sky constraint $k_\perp R=L$ or $k_\perp R=\tilde L$ given that the corrections 
to this expression are of order $\ell^{-2}$. 
This expression can be further simplified in the thin-shell
approximation and can then be further improved by including the corrections due to this thin-shell
approximation given in Eq.~(\ref{CorrectThinshell}).  Similarly to our discussion at the end of 
Section~\ref{SecIntuitiveComputation}, the first corrective term which comes
from the second term in the curly brackets in the expression above, is of order $1/\ell$
up to $\ell \simeq 100$ and then of the order of $1\%$ beyond. Including this first correction to the
lowest order of the flat-sky expansion (the first term in the curly brackets
in the expression above) considerably improves the precision of the flat-sky
expansion. Note that the corrective terms in the expression~(\ref{ClJP}) were
at least of order $1/\ell^2$, and thus the corrective term in the
expression~(\ref{ClJP2}), which comes from the directional dependence of the
sources, is the dominant one.
To compare with Eq.~(\ref{ClJP}), one would need to derive the
corrections of order $1/\ell^2$ in Eq.~(\ref{ClJP2}), which seems an unneeded
academic sophistication.
Fig.~\ref{CorrectiveTermDoppler} depicts the relative error with respect to
the exact calculation, with and without including the first correction of
order $1/\ell$ given in Eq.~(\ref{ClJP2}), using the thin-shell approximation and taking into account all effects but the late integrated Sachs-Wolfe effect.

\section{Flat-sky expansion of higher spin quantities}\label{SecPolar}

The CMB radiation is not described entirely by its temperature, since it is polarized by the
 Compton scattering of photons on free electrons. Only the linear polarization is generated through this process and it is described by the spin $\pm2$ fields defined from the Stokes parameters
\begin{equation}
_{\pm2} X(\hn) \equiv Q(\hn)\pm\ii U(\hn)\,.
\end{equation}
$_{\pm2} X(\hn)$  is dependent on the choice of the basis used to defined 
the linear polarization. Any rotation of this basis by an angle $\gamma$ around the direction $\hat \vn$ 
transforms it as $_{\pm2} X(\hn) \rightarrow _{\pm2} X(\hn)e^{ \pm 2 \ii \gamma}$. 

In order to define the correlation function of two spinned quantities, $_{s_1}X$ and $_{s_2}Y$, 
one shall use the spin raising and lowering operators, respectively $\edth^+$ and $\edth^-$ which are defined for a spin $s$ quantity by~\cite{1997PhRvD..55.1830Z,1967JMP.....8.2155G}
\begin{equation}
\edth^{\pm} {}_s X=-\sin^{\pm s}\left(\partial_\theta\pm\ii \csc \theta \partial_\varphi \right)\sin^{\mp s}{}_s X
\end{equation}
in order to relate them to a spin-0 field. We thus define
\begin{equation}
{}_s \tilde X \equiv (-1)^s \edth^{-s} {}_s X, \qquad {}_s \tilde X \equiv \edth^{-s} {}_s X 
\end{equation}
respectively for $s>0$ and $s<0$.
The expansion of ${}_s X$ on spinned spherical harmonics according to
\begin{equation}
{}_s X(\hat \vn) = \sum_{\ell m} {}_s X_{\ell m} \,{}_s Y^m_\ell(\hat \vn)
\end{equation}
can then be related to its expansion on spherical harmonics as
\begin{equation}
{}_s \tilde X(\hat \vn) = \sum_{\ell m} {}_s X_{\ell m} \,\sqrt{\frac{(\ell+|s|)!}{(\ell-|s|)!}} Y^m_\ell(\hat \vn)\,.
\end{equation}
The two sets of spherical harmonics are related by
\begin{eqnarray}
(-1)^s \edth^{-s}{}_sY^m_\ell&=&\sqrt{\frac{(\ell+s)!}{(\ell-s)!}}Y^m_\ell \quad{\rm if}\quad s>0,\\
\edth^{-s}{}_sY^m_\ell&=&\sqrt{\frac{(\ell-s)!}{(\ell+s)!}}Y^m_\ell \quad{\rm if}\quad s<0.\nonumber
\end{eqnarray}
We further define the electric and magnetic parts as
\begin{eqnarray}
E({X})(\hat \vn)&\equiv&  \frac12\left[{}_s \tilde X(\hat \vn)+{}_{-s} \tilde X(\hat \vn)\right]\,,\\
B({X})(\hat \vn)&\equiv&  \frac{1}{2 \ii}\left[{}_s \tilde X(\hat \vn)-{}_{-s} \tilde X(\hat \vn)\right]\,,
\end{eqnarray}
where here we choose the convention $s\ge 0$.
For a spin $0$ quantity such as the temperature, $E({}_0 X)={}_0 X={}_0 \tilde X$ and $B({}_0 X)=0$.
Due to parity invariance, the correlation between an electric type quantity and a magnetic type quantity always vanishes.
We then define the correlation function as
\begin{eqnarray}
&&\xi^{E\left(\tilde X\right) E\left( \tilde Y\right)}(\theta)=\langle  E(X)(\hn) \,E(Y)(\hn')^\star\rangle_{\hn.\hn'=\cos \theta}\,\\
&&\equiv\sum_{\ell} \frac{2 \ell+1}{4 \pi}\sqrt{\frac{(\ell+s_1)!}{(\ell-s_1)!}\frac{(\ell+s_2)!}{(\ell-s_2)!}} C_\ell^{E(X) E(Y)} P_{\ell}(\cos \theta)\,,\nonumber
\end{eqnarray}
with similar definitions for magnetic type multipoles.
The angular power spectra are then extracted through 
\begin{eqnarray}
C^{E\left(\tilde X\right) E\left(\tilde Y\right)}_{\ell}&=&2\pi \sqrt{\frac{(\ell-s_1)!}{(\ell+s_1)!}\frac{(\ell-s_2)!}{(\ell+s_2)!}}\\
&\times&\int\sin\theta\dd\theta\,P_{\ell}(\cos\theta)\xi^{E\left(\tilde X\right) E\left(\tilde Y\right)}(\theta)\,.\label{intthetaspinned}\nonumber
\end{eqnarray}
The emitting sources for a spin $s$ quantity are expanded similarly to Eq.~(\ref{Expsources}) 
but with a decomposition on  spinned spherical harmonics
\begin{equation}
 w[{}_s X](\vx)=\int  \frac{\dd^3 \vk}{(2 \pi)^{3/2}}  w[{}_s X](\vk,\hat \vn,r) \exp(\ii \vk.\vx)
\end{equation}
with
\begin{eqnarray}
w[{}_s X](\vk,\hat \vn,r) &=& \sum_{\ell,m} w_{\ell m}[{}_{-s} X](k,r) (\ii^\ell) \sqrt{\frac{4 \pi}{2 \ell+1}} \nonumber\\
&&\times {}_{s}Y_{\vk}^{\ell m}(\hat \vn)e^{\ii s \varphi(\vk,\hat \vn)}\,,
\end{eqnarray}
where $\varphi(\vk,\hat \vn)$ is the azimuthal angle of $\vk$ with respect to $\hat \vn$~\cite{1997PhRvD..55.1830Z}.
The source multipoles for spinned quantities are defined using the same conventions as in
Refs.~\cite{1997PhRvD..56..596H,2009CQGra..26f5006P} except that the
multipoles here refer to the direction of observation whereas in these
references it refers to the direction of propagation. We have thus made the
replacement $s \to -s$ additionally to the extra factor $(-1)^\ell$ which was already considered for spin $0$ quantities in
Eq.~(\ref{Expsources}), in order to take this fact into account by using the transformation properties under parity of the multipoles.

At the lowest order of the flat-sky expansion, we can approximate $\varphi(\vk,\hat \vn)$ by $\beta$. 
In particular, this implies that the spin raising and lowering operators applied on $w[{}_s X]$ will act only on $\exp(\ii \vk.\vx)$ and we find that at lowest order in the flat-sky expansion
\begin{equation}
\edth^{\pm s} \exp(\ii \vk.\vx)\simeq(-\ii k r)^{s} e^{\mp\ii s (\varphi-\beta)}\exp(\ii \vk.\vx),
\end{equation}
for $s>0$. We thus deduce that the sources for $E({}_s X)$ and $B({}_s X)$
are given by
\begin{eqnarray}
&&w[(E/B)({}_s X)](\vk,\hat \vn,r) =  \sum_{\ell,m} (\ii k r)^{s} (\ii)^\ell \sqrt{\frac{4 \pi}{2 \ell+1}} \\
&& \left\{\pm\frac12w_{\ell m}[(E/B)(X)](k,r)\right.\nonumber\\
&&\qquad\qquad\times\left[(-1)^s e^{-\ii s \varphi}{}_{-s}Y_{\vk}^{\ell m}(\hat \vn)+e^{\ii s \varphi}{}_{s}Y_{\vk}^{\ell m}(\hat \vn) \right]\nonumber\\
&&\quad +\frac{\ii}{2}w_{\ell m}[(B/E)(X)](k,r)\nonumber\\
&&\qquad\qquad\times\left.\left[(-1)^s e^{-\ii s \varphi}{}_{-s}Y_{\vk}^{\ell m}(\hat \vn)-e^{\ii s \varphi}{}_{s}Y_{\vk}^{\ell m}(\hat \vn) \right]\right\}\,,\nonumber
\end{eqnarray}
with the $+$ sign for $E$ and the $-$ sign for $B$. 

\begin{figure*}[!htb]
\includegraphics[width=8.5cm]{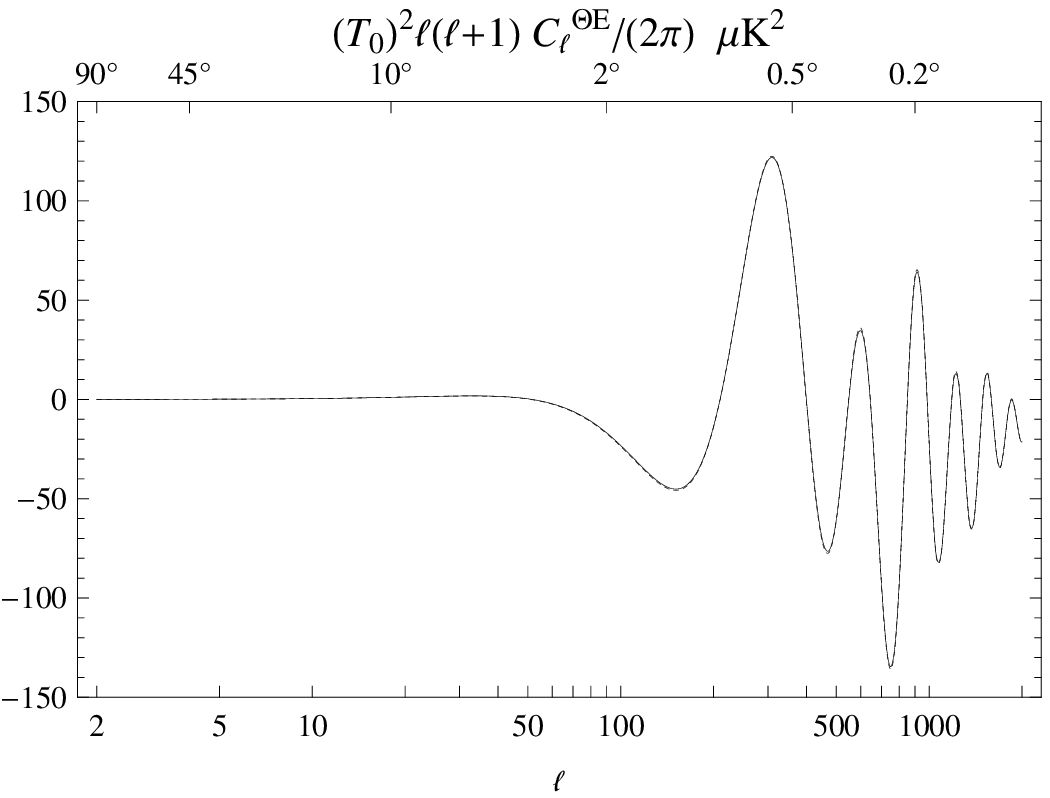}
\includegraphics[width=8.5cm]{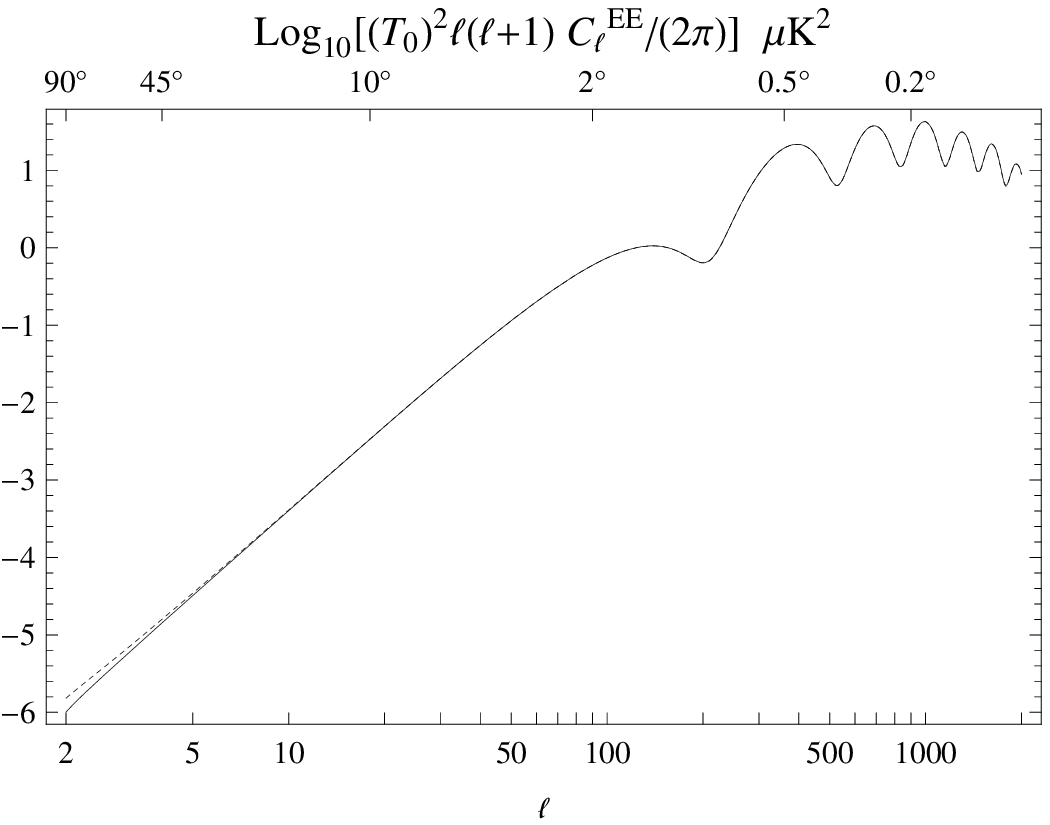}
\caption{Comparison of the flat-sky approximation (dashed line)
to the exact computation (solid line) for $C_\ell^{\Theta E}$ (left) and $C_\ell^{EE}$ (right) ignoring the effect of reionization. Since they only disagree on very large scales, in the regime where the polarization fails to be generated, both curves are hardly distinguishible. We assume standard cosmology.
}\label{TEandEE}
\end{figure*}

In the case where there are only scalar sources ($m=0$), this simplifies substantially.
From the parameterization (\ref{FSparameterizationJP}), that is with the choice $\varphi=0$ and using that ${}_sY^{\ell 0} = (-1)^s Y^{\ell s}$, we obtain
\begin{eqnarray}
&&w[E(X)](\vk,\hat \vn,r) =  \\
&&\sum_{\ell} (-\ii k r)^{s} (\ii)^\ell w_{\ell}[E(X)](k,r) \sqrt{\frac{(\ell-s)!}{(\ell+s)!}}P^{s}_\ell(k_r/k)\,\nonumber
\end{eqnarray}
and a vanishing $w[B(X)]$. 
Following the same method as in section~\ref{SecIntuitiveComputation}, the factors $(k r)^{(s_1+s_2)}$ are going to be approximately canceled by the prefactor of Eq.~(\ref{intthetaspinned}) which is behaving as $\ell^{-(s_1+s_2)}$. A more careful derivation would require to use $k=\sqrt{k_\perp^2+k_r^2}$ and the exact form of the prefactor in Eq.~(\ref{intthetaspinned}). However, we are here interested in the most simple expression for the lowest order of the flat-sky expansion and we drop these extra complications. Since polarization is not generated on large scales, this will be sufficient to obtain an excellent flat-sky expansion. In the end of the computation, for scalar perturbations the multipoles associated with the correlation of spinned quantities is obtained just by replacing the sources according to
\begin{equation}
w(k,r)\to \sum_{\ell} (-1)^s\ii^{\ell+s} w_{\ell 0}[E](k,r) \sqrt{\frac{(\ell-s)!}{(\ell+s)!}}P^s_\ell\left(\frac{k_r}{k}\right).\nonumber
\end{equation}
We use this expression for the computation of the sources of linear polarization ($s=2$, and only $\ell=2$ in the sum above since there are only quadrupolar sources), and we compare the full-sky computation of $C_\ell^{EE}$ and $C_\ell^{\Theta E}$ with the flat-sky result in Fig.~\ref{TEandEE}

\section{The bispectrum}\label{SecBispectre}

We shall now investigate the flat-sky expansion of the bispectrum. We shall assume
that it arises from primordial non-Gaussian initial
conditions described by a primordial bispectrum for the metric fluctuation, e.g., 
\begin{equation}
\langle \Phi(\vk_{1})\Phi(\vk_{2})\Phi(\vk_{3})\rangle=\delta^3(\vk_{1}+\vk_{2}+\vk_{3})\,B(\vk_{1},\vk_{2})\label{DefBprimordial}\,.
\end{equation}
As we shall see the derivation of the temperature bispectrum is much more 
subtle and we only partially succeeded in a sense we explain below.
The reason lies in the fact that there are actually many ways of representing a bispectrum
the properties of which in the flat-sky limit might be different.

\subsection{Two different representations of the bispectrum in harmonic space}

\subsubsection{Definitions}

We are interested in the expectation values of $\langle a_{\ell_{1}m_{1}}a_{\ell_{2}m_{2}}a_{\ell_{3}m_{3}}\rangle$
of the $a_{\ell m}$ coefficients of the CMB temperature. Because of the statistical isotropy of the sky, 
their $m$-dependence is bound to be that of the Gaunt integral $\mG_{m_{1}m_{2}m_{3}}^{\ell_{1}\ell_{2}\ell_{3}}$
so that it is more fruitful to introduce the reduced bispectrum 
$b_{\ell_{1}\ell_{2}\ell_{3}}$ (see Ref.~\cite{2002astro.ph..6039K} for instance) defined as
\begin{equation}
\langle a_{\ell_{1}m_{1}}a_{\ell_{2}m_{2}}a_{\ell_{3}m_{3}}\rangle\equiv \mG_{m_{1}m_{2}m_{3}}^{\ell_{1}\ell_{2}\ell_{3}}\ b_{\ell_{1}\ell_{2}\ell_{3}}.
\end{equation}
The a priori purpose of the following is to propose a controlled approximation 
for $b_{\ell_{1}\ell_{2}\ell_{3}}$ in the flat-sky limit.
It turns out however, as we shall see later, that in order to have a controlled limit expression 
of that quantity, strong regularity conditions should be imposed on the initial metric perturbation $B(\vk_{1},\vk_{2})$. 

The bispectrum can actually be equally characterized by the following quantities (reminding $L\equiv \ell+1/2$),
\begin{equation}\label{xienfonctiondeblll}
\xi_{\ell_{1}\ell_{2}M}=\sum_{\ell_{3}}{2 L_3}{\wtroisj{\ell_{1}}{\ell_{2}}{\ell_{3}}{M}{-M}{0}}{\wtroisj{\ell_{1}}{\ell_{2}}{\ell_{3}}{0}{0}{0}}\ b_{\ell_{1}\ell_{2}\ell_{3}}\,,
\end{equation}
that contains the same information as $b_{\ell_{1}\ell_{2}\ell_{3}}$ since it can be inverted as
\begin{equation}\label{xienfonctiondeblllInvert}
b_{\ell_{1}\ell_{2}\ell_{3}}=\sum_{M}\frac{\wtroisj{\ell_{1}}{\ell_{2}}{\ell_{3}}{M}{-M}{0}}{\wtroisj{\ell_{1}}{\ell_{2}}{\ell_{3}}{0}{0}{0}}\xi_{\ell_1 \ell_{2}M}.
\end{equation}
$\xi_{\ell_{1}\ell_{2}M}$ is a real-valued quantity that obeys the following symmetry properties
\begin{equation}
\xi_{\ell_{1}\ell_{2}M}=\xi_{\ell_{1}\ell_{2}-M}=\xi_{\ell_{2}\ell_{1}M}.
\end{equation}
As we shall see, $\xi_{\ell_1 \ell_{2}M}$ actually enjoys a better-controlled asymptotic expression
than $b_{\ell_{1}\ell_{2}\ell_{3}}$ in the large $\ell$ limit, similar to what was achieved
for the angular power spectrum in the previous sections.

\subsubsection{Properties of $\xi_{\ell_1 \ell_{2}M}$}

Let us first relate $\xi_{\ell_1 \ell_{2}M}$ to the angular three-point function. We introduce
three unit vectors on the celestial sphere, $\hn_{1}$, $\hn_{2}$ and $\hn_{3}$, by
\begin{equation}
\hn_{i}=(\sin\theta_{i}\cos\varphi_{i},\sin\theta_{i}\sin\varphi_{i},\cos\theta_{i}),
\end{equation}
where $\theta_{i}$ and $\varphi_{i}$ are the Euler angles and $i=1..3$.
Taking advantage of the statistical isotropy of the sky, the three-point temperature correlation 
function can always be expressed as a function of the relative angle with respect to say $\hn_{3}$, 
setting $\theta_{3}=0$, so as a function of the four angles,  $(\theta_{1},\varphi_{1},\theta_{2},\varphi_{2})$.
These dependencies can then be expanded on spherical harmonics as
\begin{eqnarray}
\langle\Theta (\hn_{1})\Theta (\hn_{2})\Theta(\hn_{3})\rangle
&\equiv& \xi(\theta_{1},\varphi_{1},\theta_{2},\varphi_{2})\\
&&\hspace{-3.5cm}=\sum_{\ell_{i},m_{i}}\xi_{\ell_{1}\ell_{2}m_{1}m_{2}}
\sqrt{\frac{2L_1 }{4\pi}\frac{2L_2}{4\pi}} Y_{\ell_{1}}^{m_{1}}(\theta_{1},\varphi_{1}) Y_{\ell_{2}}^{m_{2}}(\theta_{2},\varphi_{2})\,\nonumber.
\end{eqnarray}
Rotational invariance further implies that $\xi$ depends only on $\varphi_{21}\equiv\varphi_{2}-\varphi_{1}$ so that
$\xi_{\ell'_{1}\ell'_{2}m'_{1}m'_{2}}$ vanishes if $m'_{1}\ne -m'_{2}$. We conclude that
\begin{eqnarray}
\xi(\theta_{1},\varphi_{1},\theta_{2},\varphi_{2})=\\
&&\hspace{-3.5cm}\sum_{\ell_{1},\ell_{2},M}\xi_{\ell_{1}\ell_{2}M-M}
\sqrt{\frac{2L_1 }{4\pi}\frac{2L_2}{4\pi}} Y_{\ell'_{1}}^{M}(\theta_{1},\varphi_{1}) Y_{\ell'_{2}}^{-M}(\theta_{2},\varphi_{2})\nonumber\,.
\end{eqnarray}
This expression  generalizes the expansion of the two-point correlation function in terms of $P_{\ell}(\theta)$. 

\subsubsection{Relation between $\xi_{\ell_1 \ell_{2}M}$ and the bispectrum}

To obtain such a relation, we need  to express $\xi$ in terms of directions $\hn_{1}$, $\hn_{2}$, $\hn_{3}$ 
instead of relative angles. This can be achieved by performing a rotation $\mR$, under which the sperical
harmonics transform as
\begin{equation}
Y_{\ell}^{m}(\mR^{-1}\hn)=\sum_{m'} Y_{\ell}^{m'}(\hn)\,D^{\ell}_{m'm}(\mR)\,
\end{equation}
where $D^{\ell}_{m'm}$ are the rotation matrices and can be expressed in terms
of spin-weighted spherical harmonics as
\begin{equation}
D^{\ell}_{-ms}(\varphi,\theta,\psi)=(-1)^{m}\sqrt\frac{4\pi}{2 L } \,_{s}\!Y_{\ell}^{m}(\theta,\varphi)\,e^{-\ii s\psi}.
\end{equation}
It follows that
 \begin{eqnarray}
&&\xi(\hn_{1},\hn_{2},\hn_{3})=\sum_{\ell_{1},\ell_{2},M,m_{1},m_{2}}\xi_{\ell_{1}\ell_{2}M-M}\\
&& \times 
Y_{\ell_{1}}^{m_{1}*}(\hn_{1})\ \,_{M}\!Y_{\ell_1}^{m_{1}}(\hn_{3})
\ Y_{\ell_{2}}^{m_{2}*}(\hn_{2})\ \,_{-M}\!Y_{\ell_2}^{m_{2}}(\hn_{3}).\nonumber
\end{eqnarray}
Using
\begin{eqnarray}
&&\int{\dd^2\hn}Ê\ {\,_{s_{1}}Y_{\ell_{1}}^{m_{1}}}(\hn)\,
\,_{s_{2}}Y_{\ell_{2}}^{m_{2}}(\hn)\,
\,_{s_{3}}Y_{\ell_{3}}^{m_{3}}(\hn)=\\
&&\sqrt{\frac{8 L_1 L_2 L_3}{4 \pi}}
\wtroisj{\ell_{1}}{\ell_{2}}{\ell_{3}}{s_{1}}{s_{2}}{s_{3}}\,
\wtroisj{\ell_{1}}{\ell_{2}}{\ell_{3}}{m_{1}}{m_{2}}{m_{3}}\nonumber
\label{sGauntExp}
\end{eqnarray}
when $s_{1}+s_{2}+s_{3}=0$, we obtain
\begin{equation}\label{blllfonctiondexi}
b_{\ell_{1},\ell_{2},\ell_{3}}=\sum_{M}\frac{\wtroisj{\ell_{1}}{\ell_{2}}{\ell_{3}}{M}{-M}{0}}{\wtroisj{\ell_{1}}{\ell_{2}}{\ell_{3}}{0}{0}{0}}\xi_{\ell_1 \ell_{2}M-M}.
\end{equation}
This shows that the coefficients $\xi_{\ell_1 \ell_{2}M-M}$ are nothing but the parameters $\xi_{\ell_1 \ell_{2}M}$ 
that we introduced earlier as an alternative description of the bispectrum,
\begin{equation}
\xi_{\ell_1 \ell_{2}M}=\xi_{\ell_1 \ell_{2}M-M}\,.
\end{equation}
They can therefore be expressed in terms of the real space correlation function,
\begin{eqnarray}\label{intFullSky4}
\xi_{\ell_1 \ell_{2}M}&=&\frac{4 \pi}{\sqrt{2 L_1 2 L_2}}\int\dd^2\hn_{1}\dd^2\hn_{2}
\xi(\theta_{1},\varphi_{1},\theta_{2},\varphi_{2})\nonumber\\
&& \times Y_{\ell_{1}}^{M}(\theta_{1},\varphi_1)\,Y_{\ell_{2}}^{-M}(\theta_{2},\varphi_{2})\\
&=&\frac{8 \pi^2}{\sqrt{2 L_1 2 L_2}} \int\sin\theta_{1}\dd\theta_{1}\int\sin\theta_{2}\dd\theta_{2} \int \dd\varphi_{21}\ 
\nonumber\\
&&\times Y_{\ell_{1}}^{M}(\theta_{1},0)\,Y_{\ell_{2}}^{-M}(\theta_{2},\varphi_{21})\ \xi(\theta_{1},0,\theta_{2},\varphi_{21}).\nonumber
\end{eqnarray}
This is the generalisation of Eq.~(\ref{inttheta}) for the bispectrum.

\subsection{Flat-sky limit of $\xi_{\ell_1 \ell_{2}M}$}

We follow the same path as for the power spectrum. The first step is then to provide a formal expression of the three-point correlation function in real space. Setting
\begin{equation}
\vx_{i}=r_{i}\hn_{i},
\end{equation}
for $i=1...3$ and
\begin{eqnarray}
\vk_{1}&=&(k_{1}\sin\alpha_{1}\cos\beta_{1},k_{1}\sin\alpha_{1}\sin\beta_{1},k_{1}\cos\alpha_{1})\nonumber\\
&=&
(k_{1}^{\perp}\cos\beta_{1},k_{1}^{\perp}\sin\beta_{1},k_{1}^z)
,\\ 
\vk_{2}&=&(k_{2}^{\perp}\cos\beta_{2},k_{2}^{\perp}\sin\beta_{2},k_{2}^{z})\,.
\end{eqnarray}
Formally the three-point temperature correlation function then reads
\begin{widetext}
\begin{eqnarray}
\xi(\theta_{1},\varphi_{1},\theta_{2},\varphi_{2})&=&\int\dd^3\vk_{1}\dd^3\,\vk_{2}\,\dd r_{1}\,\dd r_{2}\,\dd r_{3}\,
B(\vk_{1},\vk_{2})\exp\left[\ii k_{1}^{z}(r_{1}\cos\theta_{1}-r_{3})+\ii k_{2}^{z}(r_{2}\cos\theta_{2}-r_{3})\right]\nonumber\\
&&\times w({k_{1}},r_{1})\,w({k_{2}},r_{2}) w(\vert\vk_{3}\vert,r_{3})\exp\left[\ii k_{1}^{\perp}r_{1}\sin\theta_{1}\cos(\beta_{1}-\varphi_{1})+\ii k_{2}^{\perp}r_{2}\sin\theta_{2}\cos(\beta_{2}-\varphi_{2})\right]\,,\label{xi3express}
\end{eqnarray}
\end{widetext}
where the Dirac distribution of Eq.~(\ref{DefBprimordial}) has been taken into account. We are left with a function of the relative angles $(\theta_{1},\varphi_{1})$
and $(\theta_{2},\varphi_{2})$.

For a fixed value of $M$, $Y_{\ell}^{M}(\theta,\varphi)$ has a well controlled limit in the flat-sky approximation. It is given by Eq.~(8.722) of Ref.~\cite{GR}
\begin{equation}\label{EqYlM}
Y_{\ell}^{M}(\theta,\varphi)\to \left(\frac{2 L}{4\pi}\right)^{1/2}(-1)^{M}J_{M}[L\theta]e^{\ii M \varphi}\,.
\end{equation}
In appendix \ref{AppYlmAsymptotics}, we show how the next to leading order terms of this expression can be obtained. The expansion parameter is $M/\ell$ or $M\theta$.
The completion of the calculation then relies on the relation,
\begin{equation}\label{Orthom}
\int_0^\infty x J_M(a x)J_M(b x) \dd x = \frac{\delta(a-b)}{b}\,.
\end{equation}
We can now proceed to the evaluation of (\ref{intFullSky4}) in the flat-sky limit, as it is now straightforward. 
Defining $\rho_{1}=k_{1}^{\perp}r_{1}\theta_{1}$ and $\rho_{2}=k_{2}^{\perp}r_{2}\theta_{2}$
the expression of $\xi(\theta_{1},\varphi_{1},\theta_{2},\varphi_{2})$ at leading order is
\begin{eqnarray}\label{xi3smalla}
&&\xi(\theta_{1},0,\theta_{2},\varphi_{21})=\\
&&\int\dd^3\vk_{1}\dd^3\,\vk_{2}\,\dd r_{1}\,\dd r_{2}\,\dd r_{3}\,B(\vk_{1},\vk_{2})\left[\prod_{a=1,2,3}w(k_{a},r_{a})\right] \nonumber\\
&&\exp\left[\ii \sum_{a=1,2}k_{a}^{z}r_{a3}\right] \exp\left[\ii \rho_{1}\cos\beta_{1}+\ii \rho_{2}\cos(\beta_{2}-\varphi_{21})\right]\,,\nonumber
\end{eqnarray}
with $r_{i3}\equiv r_i-r_3$, and $k_3 \equiv \vert \vk_3\vert$.
Then the subsequent angular integrations that appear in the expression of $\xi_{\ell_1 \ell_{2}M}$ 
in Eq.~(\ref{intFullSky4}) lead to the following transforms:
\begin{itemize}
\item the integration over $\varphi_{21}$ of  $\exp\left[\ii \rho_{2}\cos(\beta_{2}-\varphi_{1})-\ii M\varphi_{21}\right]$
gives a term  $\ii^{M}J_{M}(\rho_{2})\,e^{\ii M\beta_{2}}$;
\item the integration over $\beta_{1}$, fixing the relative angle $\beta_{12}$ between 
the wave vectors $\vk_{1}$ and $\vk_{2}$, of  $\exp\left[\ii \rho_{1}\cos\beta_{1}+\ii M\beta_{2}\right]$
gives a term $(-\ii)^MJ_{M}(\rho_{1})\,e^{\ii M\beta_{12}}$;
\item the integration over $\theta_{1}$ of $J_{M}(\rho_{1})J_{M}(L_1\theta_{1})$
gives a term  $\delta(L_1-k_{1}^{\perp}r_{1})/L_1$;
\item the integration over $\theta_{2}$ of $J_{M}(\rho_{2})J_{M}(L_2\theta_{2})$
gives a term  $\delta(L_2-k_{2}^{\perp}r_{2})/L_2$.
\end{itemize}
Then, the integration over $k_{1}^{\perp}$ and $k_{2}^{\perp}$ can be performed explicitly and
we are left with
\begin{equation}
\xi_{\ell_1 \ell_{2}M}=
\int_{0}^{2\pi}\frac{\dd \beta_{\{k\}}}{2\pi}\ e^{\ii M\beta_{\{k\}}}\ b^{\rm fs}_{\ell_{1}\ell_{2}}(\beta_{\{k\}})
\end{equation}
where
\begin{eqnarray}
&&b^{\rm fs}_{\ell_{1}\ell_{2}}(\beta_{\{k\}})=\int\dd k_{1}^{z}\,\dd k_{2}^{z}\,\dd r_{1}\,\dd r_{2}\,\dd r_{3}\,
\frac{w(k_{1},r_{1})}{r_{1}^2}\frac{w(k_{2},r_{2})}{r_{2}^2}\nonumber\\
&& \qquad \qquad \times w(k_3,r_{3})B(\vk_{1},\vk_{2})\exp\left[\ii k_{1}^{z}r_{13}+\ii k_{2}^{z}r_{23}\right].\label{bl1l2fs}
\end{eqnarray}
In this expression,
\begin{eqnarray}
k_{1}^2&=&{k_{1}^z}^2+L_1^2/r_{1}^2,\qquad
k_{2}^2={k_{2}^z}^2+L_2^2/r_{2}^2,
\end{eqnarray}
and $B(\vk_{1},\vk_{2})$ is an implicit relation of $k_{1}^z$, $k_{2}^z$ and of the relative angle 
of their transverse parts, $\beta_{\{k\}}$ since
\begin{eqnarray}
k_{3}^2&=&(k_{1}^z+k_{2}^z)^2+L_1^2/r_{1}^2+L_2^2/r_{2}^2\nonumber\\
&&+2\cos(\beta_{\{k\}})L_1 L_2/(r_{1}r_{2}).
\end{eqnarray}
It can further be noted that because $b^{\rm fs}_{\ell_{1}\ell_{2}}(\beta_{\{k\}})$ is invariant 
under $\beta_{\{k\}}\to-\beta_{\{k\}}$, the expression of $\xi_{\ell_1 \ell_{2}M}$ also reads
\begin{equation}
\xi_{\ell_1 \ell_{2}M}=
\int_{0}^{2\pi}\frac{\dd \beta_{\{k\}}}{2\pi}\ \cos(M\beta_{\{k\}})\ b^{\rm fs}_{\ell_{1}\ell_{2}}(\beta_{\{k\}}).
\label{xil1l2Mfs}
\end{equation}
The $M$-dependence in this expression is the one of the Fourier transform 
of the $\beta_{\{k\}}$-dependence, that is that of the relative angle between the wave 
numbers in the transverse direction. The equations (\ref{bl1l2fs})
and (\ref{xil1l2Mfs}) represent the flat-sky approximation of
$\xi_{\ell_1 \ell_{2}M}$.

\subsection{The flat-sky limit of $b_{\ell_1 \ell_{2}\ell_{3}}$}

While it was straightforward to derive the flat-sky limit of $\xi_{\ell_1 \ell_{2}M}$, the one
of $b_{\ell_1 \ell_{2}\ell_{3}}$ is more problematic. First, $b_{\ell_1 \ell_{2}\ell_{3}}$
is obtained from a sum of contributions each of which involves
$\xi_{\ell_1\ell_{2}M}$ which should be calculated in the flat-sky limit. If
the number of $M$ in that sum is finite, it is however possible to obtain  $b_{\ell_1 \ell_{2}\ell_{3}}$.
There is however a priori no reasons for the sum in Eq.~(\ref{blllfonctiondexi}) to be dominated 
by its first terms. It depends actually on the details of the model and in particular 
on the regularity of $B(\vk_1,\vk_2)$. Fortunately, this should be the
case for the models of interest in cosmology.

We can then try to invert Eq.~(\ref{xienfonctiondeblll}) in the large $\ell_3$ limit. 
First, Eqs.~(6.578.8) and (8.754.2) of Ref.~\cite{GR} (which are a particular case 
of the Ponzano and Regge semiclassical limit of the Wigner coefficients~\cite{PonzanoRegge68})
allow to infer the limit
\begin{eqnarray}
\frac{\wtroisj{\ell_{1}}{\ell_{2}}{\ell_{3}}{M}{-M}{0}}{\wtroisj{\ell_{1}}{\ell_{2}}{\ell_{3}}{0}{0}{0}}
&\to&
\frac{\int\theta\dd\theta\ J_{M}(\ell_{1}\theta)J_{M}(\ell_{2}\theta)J_{0}(\ell_{3}\theta)}{\int\theta\dd\theta\ J_{0}(\ell_{1}\theta)J_{0}(\ell_{2}\theta)J_{0}(\ell_{3}\theta)}\nonumber\\
&=&\frac{P_{M-1/2}^{1/2}(\cos\gamma_{12})}{P_{-1/2}^{1/2}(\cos\gamma_{12})}=\cos M\gamma_{12}
\end{eqnarray}
where $\gamma_{12}$ is the angle between ${\bf L}_{1}$ and ${\bf L}_{2}$ if $(L_{1},L_{2},L_{3})$ forms a triangle. Such a limit is 
valid again for finite values of $M$ only. 

It is then possible to transform the discrete sum on $\ell_3$ in
Eq.~(\ref{xienfonctiondeblll}) into a continuous integral on $\ell_3$. From
the expression of the Ponzano-Regge limit

\begin{equation}
{\wtroisj{\ell_{1}}{\ell_{2}}{\ell_{3}}{0}{0}{0}}^2=\frac{1}{2\pi {{\cal A}(L_1,L_2,L_3)}}
\end{equation}
where $ {{\cal A}(L_1,L_2,L_3)}$ is the area of the triangle formed by $L_1$, $L_2$ and $L_3$, we can easily obtain that~\footnote{Note that because $\ell_{1}+\ell_{2}+\ell_{3}$ should be even, subsequent values of $\ell_{2}$ in this sum are separated by 2 units.}
\begin{eqnarray}
\xi_{\ell_{1}\ell_{2}M}&=&\sum_{\ell_{3}}2L_3{\wtroisj{\ell_{1}}{\ell_{2}}{\ell_{3}}{M}{-M}{0}} {\wtroisj{\ell_{1}}{\ell_{2}}{\ell_{3}}{0}{0}{0}}\,b_{\ell_{1}\ell_{2}\ell_{3}}\nonumber\\
&&\to\int\frac{\dd\beta_{\{L\}}}{2\pi}\cos(M\beta_{\{L\}})\ b_{\ell_{1}\ell_{2}\ell_{3}}
\end{eqnarray}
where $\beta_{\{L\}}$ is the angle formed by $L_{1}$ and $L_{2}$, i.e.
\begin{equation}
L_3^2=L_1^2+L_2^2+2L_1 L_2\cos\beta_{\{L\}}.
\end{equation}
It has to be emphasized that this continuous limit can only be taken 
when not only $\ell_{1}$ and $\ell_{2}$ are large but also when their
difference $\vert\ell_{1}-\ell_{2}\vert$ is large, so that the sum is not dominated 
by discrete values when $\beta_{\{L\}}\to \pi$. Given this limitation one gets by identification the expression 
of the $b_{\ell_{1},\ell_{2},\ell_{3}}$ in the flat-sky limit
\begin{equation}
b_{\ell_{1}\ell_{2}\ell_{3}}^{\rm fs}=b^{\rm fs}_{\ell_{1}\ell_{2}}[\beta_{\{L\}}]\\.\label{FormuleFlatsky}
\end{equation}
This means that it fixes $\beta_{\{k\}}=\beta_{\{L\}}$ and thus fixes the value 
of $k_3^\perp$ as a function of $r_3$ and $L_3$. This relation is actually nontrivial 
as it is not given by $k^\perp_{3}=L_{3}/r_{3}$ as one would naively expect. It is instead given by
\begin{equation}\label{k3perpExression}
{k_{3}^{\perp}}^2=\frac{L_{3}^2}{r_{1}r_{2}}+\frac{L_{1}^2}{r_{1}r_{2}}\left(\frac{r_{2}}{r_{1}}-1\right)+\frac{L_{2}^2}{r_{1}r_{2}}\left(\frac{r_{1}}{r_{2}}-1\right).
\end{equation}
If the last scattering is thin enough so that we can approximate $r_1=r_2=r_3$, that is in the thin 
shell approximation, we recover what we would have naively expected, i.e. $k_3^\perp r_3 =L_3$.

Similarly to the angular power spectrum case, we have used $L=\ell+1/2$ instead of $\ell$ to 
improve the convergence of the flat-sky expansion. 
We have not been able however to infer formally the validity of such an expression. In other words, 
we were not able to compute the next to leading order terms of this expression. 

\begin{figure*}[!htb]
\includegraphics[width=8.5cm]{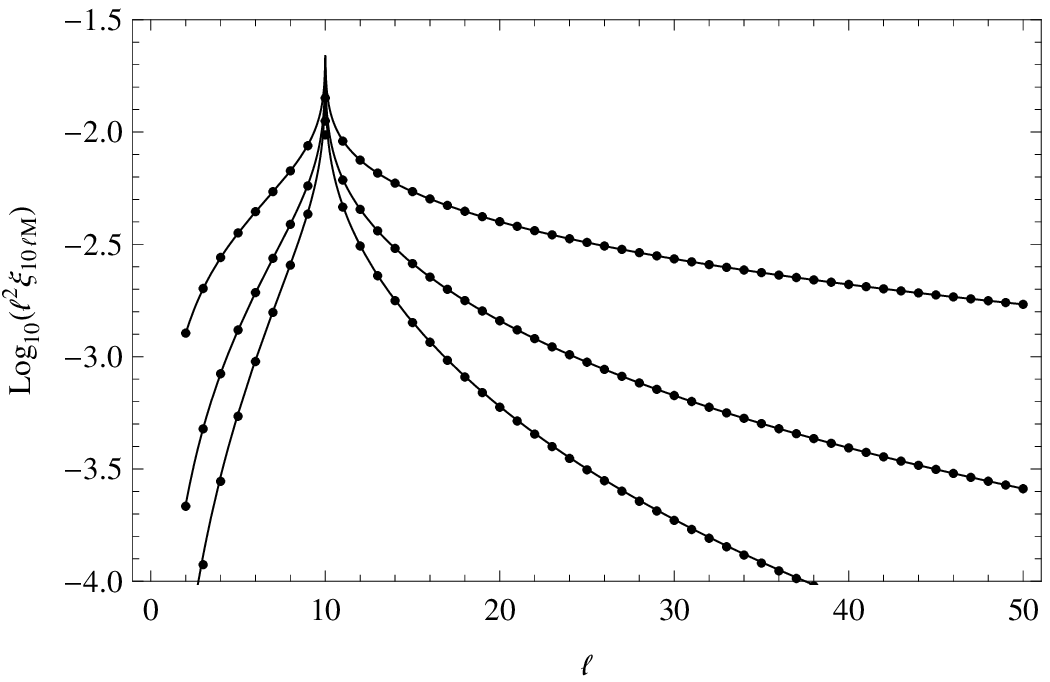}
\includegraphics[width=8.5cm]{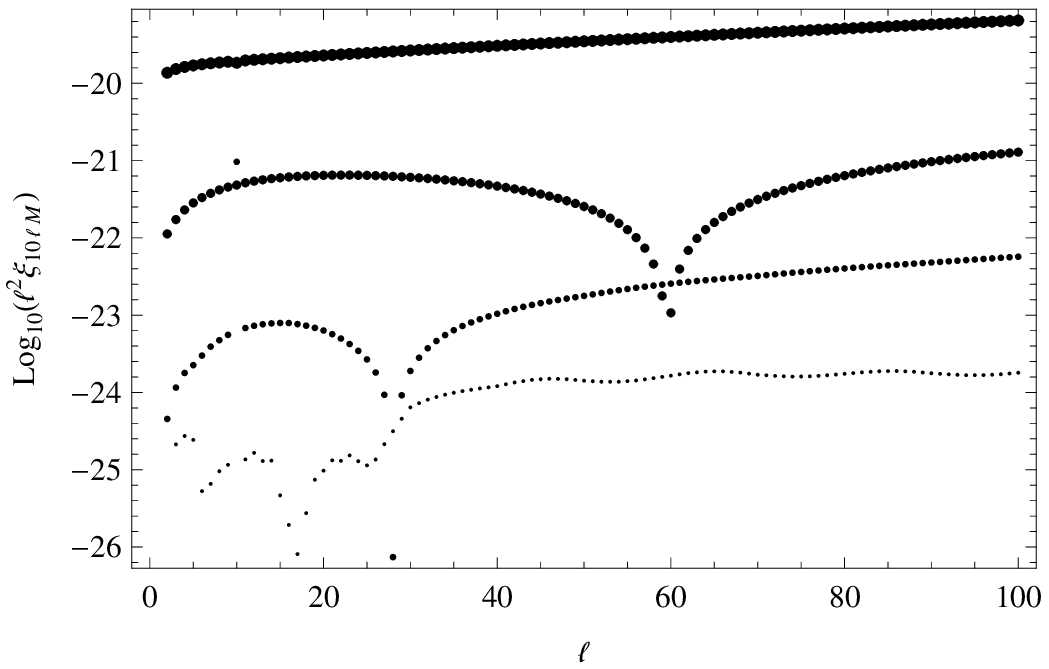}
\caption{{\em Left:} $\xi^{\rm flat}_{\ell_{1}\ell M}$ for
  $\ell_{1}=10$, $M=0,1,2$ (top to bottom) from exact computations
  (dots) and flat-sky approximation (solid line) in the Sachs-Wolfe
  plateau limit. {\em Right:} $\xi^{\rm local}_{\ell_{1}\ell M}$ for $\ell_{1}=10$, $M=0,1,2,3$ (from thick to thin dots) 
computed exactly from Eq.~(\ref{xienfonctiondeblll}).} \label{xiflat} \label{xilocal}
\end{figure*}

\subsection{Examples}

We explore the consequences of our flat-sky expression of the bispectrum
on a series of example in order to show that it is a robust
expression in most practical cases.

\subsubsection{Flat case}

We first consider the so-called flat case for which the primordial bispectrum takes the form
\begin{equation}
B^{\rm flat}(k_{1},k_{2},k_{3})\sim \frac{1}{k_{1}^2k_{2}^2k_{3}^2}.
\end{equation}
In the Sachs-Wolfe limit, the
bispectrum of the CMB temperature can be computed exactly and is given~\cite{Bartolo:2004if} by
\begin{equation}
b^{\rm flat}_{\ell_{1}\ell_{2}\ell_{3}}\sim\frac{2}{L_1 L_2 L_3}\left[\frac{L_1+L_2+L_3}{(L_1+L_2+L_3)^2-9/4}\right]
\end{equation}
whereas our flat-sky limit~(\ref{FormuleFlatsky}) gives
\begin{equation}
b^{\rm flat}_{\ell_{1}\ell_{2}\ell_{3}}\sim\frac{2}{L_1 L_2 L_3(L_1+L_2+L_3)}.
\end{equation}
These two expressions differ only by a term of the order of $1/(L_1+L_2+L_3)^2$. 

The computation of $\xi_{\ell_{1}\ell_{3}M}$ is more complicated since there is no simple expression
available. In the flat-sky approximation we get
\begin{equation}
\xi^{\rm flat}_{\ell_{1}\ell_{3}M}=\frac{M\pi-c_{M}E\left(\frac{4 \ell _1 \ell _3}{\left(\ell _1+\ell
   _3\right)^2}\right)-d_{M}K\left(\frac{4 \ell _1 \ell _3}{\left(\ell
   _1+\ell _3\right)^2}\right) }{\pi\ell _1^2 \ell _3^2}\label{xiMexp}
\end{equation}
where $E$ and $K$ are Elliptic functions of the first and second kinds respectively.
$c_{M}$ and $d_{M}$ are homogeneous functions of $\ell_{3}/\ell_{1}$ that are 
such that the numerator in Eq.~(\ref{xiMexp}) scales like $(\ell_{3}/\ell_{1})^{-1-M}$ 
when $\ell_{3}/\ell_{1}$ is large,
\begin{eqnarray}
c_{0}&=&1=-d_{0}\\
c_{1}&=&1=d_{1}\\
c_{2}&=&\frac{(2\ell_{3}/\ell_{1} + 1)(\ell_{3}/\ell_{1} + 2)}{\ell_{3}/\ell_{1} }\\
d_{2}&=&-
\frac{(2(\ell_{3}/\ell_{1})^2 - 3 \ell_{3}/\ell_{1} + 2)}{\ell_{3}/\ell_{1} }\,.
\end{eqnarray}
The result formally diverges for $\ell_{1}\to \ell_{3}$, due to the fact that 
some configurations are IR divergent when $k_{2}\to 0$. 
It implies that for the flat case in the Sachs-Wolfe plateau limit, the set of $M$'s 
contributing in the sum (\ref{blllfonctiondexi}) remains finite and this validates 
the flat-sky approximation for the bispectrum given by Eq.~(\ref{FormuleFlatsky}).
The result is depicted on Fig.~\ref{xiflat}.

\subsubsection{Local case}

In the local case, the bispectrum is defined~\cite{Bartolo:2004if} by
\begin{equation}
B^{\rm local}(k_{1},k_{2},k_{3})\sim \frac{1}{k_{1}^3k_{2}^3}+\frac{1}{k_{2}^3k_{3}^3}+\frac{1}{k_{3}^3k_{1}^3}.\label{Blocalprimordial}
\end{equation}
The bispectra can be computed by splitting $B$ in 3 terms none 
of which has pathological IR divergences 
(in other words, $b_{\ell_{1}\ell_{2}\ell_{3}}$ can be symmetrized only at the end) and
we shall not encounter the divergence problems we had in the flat case.
In its thin-shell approximation, a similar method consisting in splitting the computation 
in three terms was also of particular interest for the numerical computation of 
the bispectrum generated by non-linear effects around the LSS that we performed in 
Ref.~\cite{2010JCAP...07..003P}.

Furthermore, in the Sachs-Wolfe regime, the only non-zero contribution 
for each term is for $M=0$. This leads to the flat-sky limit of the bispectrum
\begin{equation}
b_{\ell_{1}\ell_{2}\ell_{3}}^{\rm local} = \left(\xi^{\rm local}_{\ell_{1}\ell_{3}0}+\xi^{\rm local}_{\ell_{2}\ell_{1}0}+\xi^{\rm local}_{\ell_{3}\ell_{2}0}\right)
\end{equation}
with
\begin{equation}
\xi^{\rm local}_{\ell_{1}\ell_{2}0}\sim\frac{1}{(L_1 L_2)^2}\,.
\end{equation}
Had we chosen the flat constraint $\tilde L=k_\perp r$ instead of $L=k_\perp r$, 
we would have obtained a similar expression with $L \to \tilde L$. 
It would thus have matched the full-sky expression, similarly to what was obtained in Section~\ref{SecRapConv}.
Again, we conclude that for the local case, in the Sachs-Wolfe plateau limit, only $M=0$ is contributing and this validates the assumption that the number of $M$'s contributing in the sum (\ref{blllfonctiondexi}) is finite. 
Beyond the Sachs-Wolfe regime we can estimate numerically the different terms in the sum on $M$ in order to validate 
the assumption that only a finite number of $M$ is required to estimate the bispectrum out of the $\xi_{\ell_1 \ell_2 M}$. Similarly to the Sachs-Wolfe plateau, we split in three terms the primordial bispectrum~(\ref{Blocalprimordial}),
and for each of these terms we plot $\xi_{\ell_1 \ell2 M}$ on Fig.~\ref{xilocal} for different values of $M$. It
is clear that the contribution of each $M$ is exponentially supressed when $M$ increases.
%

\begin{figure}[!htb]
\includegraphics[width=8.5cm]{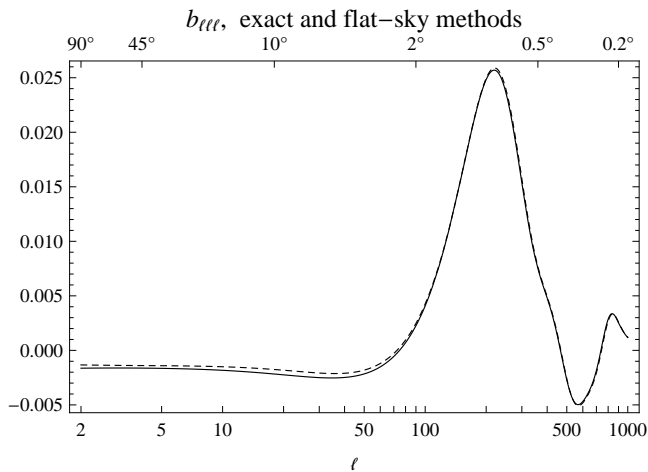}
\caption{Comparison of the flat-sky approximation (dashed line) to the exact full sky
computation (solid line) for the reduced bispectrum $b_{\ell \ell \ell}$ in the case
of local type primordial non-Gaussianity.} \label{flatskyblll}
\end{figure}

To finish, we compare in Fig.~\ref{flatskyblll} the flat-sky limit to the exact full-sky calculation 
for the bispectrum of local type and for equilateral configurations (i.e. such that $\ell_1=\ell_2=\ell_3=\ell$).
The agreement is excellent.

\section{Conclusion}

This article provides a systematic construction of the flat-sky approximation
of the angular power spectrum both for the temperature and
polarization, in particular it shows that the expansion
can be performed to any order. Additionally, we showed that
this construction is not unique and that
there exists a two-parameter family of flat-sky expansions, depending
on the arbitrary choice of the flat-sky directions with respect to the azimuthal angle.

As long as the sources are scalar, the first correction term scales as $1/\ell^2$
for a proper choice of flat-sky constraints ($k_\perp r = \ell+1/2$ or
$k_\perp r = \sqrt{\ell(\ell+1)}$). In more realistic
cases involving direction-dependent sources, such as the Doppler term or the
anisotropic stress, the corrections terms were shown to scale as $1/\ell$
whatever the flat-sky constraint and are given in Eq.~(\ref{ClJP2}). 

Two particular extensions of the flat-sky expansion are particularly useful: the thin-shell and the Limber approximations,
depending on the spatial extension of the sources. We checked that the corrective terms obtained in the Limber approximation are consistent with existing literature~\cite{2008PhRvD..78l3506L}, and we recovered the expression 
derived at leading order in the thin-shell approximation~\cite{Bond1996}
where all flat-sky expressions are identical. We discussed here the validity of those approximations.

For practical purposes, the ``best'' formula to use depends on the context:
\begin{enumerate}
 \item for the CMB, as long as the thin-shell approximation is a good
   approximation, that is on the LSS, one should use Eq.~(\ref{ClJP2}) since one cannot neglect the Doppler term.
 It includes corrections of order $1/\ell$,  and its validity is limited by the errors introduced while taking the thin-shell limit which are
 at most of order $\Delta r_\lss/r_\lss<0.01$. They appear in practice
 to be less on small scales and on large scales they can be corrected with Eq.~(\ref{CorrectThinshell}).
 \item for large scale structures such as galaxy catalogs or
   weak-lensing, for which the Limber approximation is a good approximation, one should use Eq.~(\ref{ClLimber2}).
It includes corrections of order $1/\ell^2$ and its validity is
limited to $1/\ell^3$~\footnote{Note that this differs from the
  conclusion of Ref.~\cite{2008PhRvD..78l3506L} where it is claimed without proof that
  the next order is only at $\ell^{-4}$.}, as long as there are no larger errors introduced
while taking the Limber approximation. Note also that for the late
integrated effects and the effect of reionization on the CMB, we
should also use a Limber approximation.
\end{enumerate}

Generalization of this construction scheme to the bispectra was found to be more cumbersome. It led us to introduce an alternative description of the bispectra for which the flat-sky approximation is well controlled. This corresponds to the coefficients
$\xi_{\ell_{1}\ell_{2}M}$ defined in Eq.~(\ref{xienfonctiondeblll}) whose relation with the usual reduced bispectrum form
$b_{\ell_{1}\ell_{2}\ell_{3}}$ can be found in Eq.~(\ref{xienfonctiondeblll})  and inverted in Eq.~(\ref{xienfonctiondeblllInvert}).
For this quantity we were able to propose a well controlled flat sky approximation. It actually leads to a specific
form of  flat-sky approximation  for $b_{\ell_{1}\ell_{2}\ell_{3}}$, as described in Eq.~(\ref{bl1l2fs}) in the sense given by
Eqs. (\ref{FormuleFlatsky}-\ref{k3perpExression}), 
the next-to-leading order
terms of which remain however obscure (we encounter here exactly the same difficulty as when one tries to use the correspondance
of Eq.~(\ref{corres})). In this case also one can further simplify the numerical integrations by using the thin-shell
or Limber approximations.

The validity of the bispectrum flat-sky leading order expansion
was tested numerically in simple cases of separable primordial bispectrum such as the local
and flat primordial bispectra. It was found to be very accurate in those cases, below the 1\% level.  Such expressions
are obviously of great interest for non-separable shapes of primordial non-Gaussianity 
since no fast full-sky method is known yet.  
%

\acknowledgements

C.P. is supported by STFC (UK) grant ST/H002774/1 and would like to thank Institut d'Astrophysique de Paris for its kind hospitality during part of this project.  All computations, for the spectra and bispectra, have been performed using the freely available {\it Mathematica} code which is available on the webpage~\cite{CMBquick}.

\appendix
\section{Useful properties of the Bessel functions}\label{BesselOrtho}

The Bessel functions satisfy the recursion relations for $n \ge 1$
\begin{eqnarray}
J_n'(x)&=&J_{n-1}(x)-\frac{n}{x}J_n(x)\,,\\
J_{n+1}(x)&=&\frac{2 n}{x}J_n(x)-J_{n-1}(x)\,,
\end{eqnarray}
with $J_0'=-J_1$.

We define the integrals
\begin{equation}
I_{n,p}(a,b)\equiv \int_0^\infty x^{(2p+n+1)} J_n(a x)J_0(b x) \dd x\,.
\end{equation}
By taking successive derivatives with respect to $a$ we obtain
\begin{eqnarray}
I_{1,p}(a,b)&=&-\partial_a I_{0,p}(a,b)\\
I_{0,p+1}(a,b)&=&\frac{1}{a}\partial_a \left[a I_{1,p}(a,b)\right].
\end{eqnarray}
Using the orthonormality relation
\begin{equation}\label{Ortho0}
\int_0^\infty x J_0(a x)J_0(b x) \dd x = \frac{\delta(a-b)}{b}\,,
\end{equation}
we obtain the expression of the main integrals of interest for this paper
\begin{eqnarray}
\hspace{-0.5cm}I_{0,0}(a,b)&=&\frac{\delta(b-a)}{b}\label{Ortho1}\\
\hspace{-0.5cm}I_{1,0}(a,b)&=&\frac{\delta'(b-a)}{b}\label{Ortho2}\\
\hspace{-0.5cm}I_{0,1}(a,b)&=& \frac{\delta'(b-a)}{ab}-\frac{\delta''(b-a)}{b}\label{Ortho3}\\
\hspace{-0.5cm}I_{1,1}(a,b)&=&\frac{\delta'(b-a)}{b a^2}+ \frac{\delta''(b-a)}{ab}-\frac{\delta'''(b-a)}{b}.\label{Ortho4}
\end{eqnarray}

\section{Relating the corrections in the different flat-sky expansions}\label{AppRelationsFS}

In the expansion~(\ref{Plexpan3}), had we chosen the constraint $k_\perp R=\tilde L$, then
the expression of the corrective terms in the square brackets would have been modified by the replacement of $L$
by $\tilde L$ and an extra operator $-1/(8\tilde L^2) D_\perp$ would
have been added in the brackets since  for any function $Q$,
$Q(\tilde L/r)\simeq Q(L/r)-1/(8\tilde L^2) k Q'(k)$. This means that  the numerical factor in front of the operator $D_\perp$ would be
$1/6$ instead of $1/24$. For a scale-invariant power spectrum and a sharply peaked constant transfer function,
i.e. $w(k,r)\propto\delta(r-r_\lss)$, the correction terms are thus proportional to
\begin{equation}
(4 D_\perp - D_\perp^3)\frac{1}{k^3} = 15 k_\perp^4\frac{(k_\perp^2-6 k_r^2)}{k^9}\,,
\nonumber
\end{equation}
and the integral on $k_r$ of this function is zero. This is 
a good check of our expansion. Indeed, in Section~\ref{SecRapConv} we have seen that in 
this particular case, the leading order of the flat-sky expansion is equal to 
the exact result so that it was expected that the corrections
vanish. Actually, this property should not depend on $v$  since the $v$-dependence in the lowest order term of the expansion coming from
the fact that $R$ is a function of $v$, disappears when the sources are
located only at a given distance $r_\lss$. For all choices of $v$ and for a scale invariant power spectrum and 
a peaked transfer function, the lowest order of the flat-sky expansion
is equal to the exact result. We can check indeed that
in that case the corrective terms in the expression~(\ref{ClFB}) where $v=0$ once expressed with the
constraint $k_\perp R = \tilde L$, vanish as well. The $v$-dependent
corrections need to be consistent, and though they are formally different as $v$ is varied, they should lead to the same
result, at least in that case where the transfer function is sharply
peaked. We now show that the corrective terms for different choices of
$v$ can be related by integration by parts in that case. 

We assume that $w(k,r) \propto \delta(r-r_\lss)$ so that we can 
also assume that all the $k$ dependence is located in $P(k)$ so that we can drop the factors $w(k,r_\lss)$ for simplicity and set $r_\lss=1$. 
Expanding the correlation~(\ref{xiexp2}) in $Z$ we obtain
\begin{eqnarray}
\xi_v &=& \xi_{1/2}+\int \frac{\dd k_{r}}{(2\pi)^2}\ k_{\perp}\dd k_{\perp} P(k)\\
&&\times \frac{(1-2u)Z^3 }{8}\left[-Z k_r^2J_0(k_\perp Z)+k_\perp J_1(k_\perp Z)  \right]\nonumber
\end{eqnarray}
and thus we can relate the corresponding multipoles
\begin{equation}
C_\ell^v=C_\ell^{(v=1/2)}+ \Delta C_\ell\,,
\end{equation}
with
\begin{eqnarray}
\Delta C_\ell &=& \frac{(1-2v)}{8}\int \frac{\dd k_{r}}{(2\pi)}\ \dd k_{\perp} P(k)\\
&&\times\left[k^2_\perp I_{1,1}(k_\perp,\tilde L) - k_\perp k_r^2 I_{0,2}(k_\perp,\tilde L)\right]\nonumber\\
&=& \frac{(1-2v)}{8}\int \frac{\dd k_{r}}{(2\pi)}\ \dd k_{\perp} P(k)\nonumber\\
&&\times\left\{k^2_\perp I_{1,1}(k_\perp,\tilde L) - k_r^2\frac{\partial}{\partial k_\perp}\left[k_\perp I_{1,1}(k_\perp,\tilde L)\right]\right\}.\nonumber
\end{eqnarray}
Integrating by parts we obtain
\begin{eqnarray}
\Delta C_\ell&=& \frac{(1-2v)}{8}\int \frac{\dd k_{r}}{(2\pi)}\ \dd k_{\perp}\\
&&\times k^2_\perp\left\{ P(k)I_{1,1}(k_\perp,\tilde L) + \frac{k_r^2}{k} P'(k) I_{1,1}(k_\perp,\tilde L)\right\}\,.\nonumber
\end{eqnarray}
Another integration by parts in $k_r$ is sufficient to obtain that $\Delta C_\ell = 0$. 
This proves that even if the expression of the corrections is formally different for a different $v$, 
the corrective term remains the same in the thin-shell approximation.

\section{Asymptotic forms for the spherical harmonics}\label{AppYlmAsymptotics}

The spherical harmonics can be expressed in terms of the Legendre polynomials as
\begin{equation}
Y_{\ell}^{m}(u,\varphi)=\left(\frac{2\ell+1}{4\pi}\right)^{1/2}\left[\frac{(\ell-m)!}{(\ell+m)!}\right]^{1/2}\,P_{\ell}^{m}(u)\,e^{\ii m\varphi}
\end{equation}
where $u=\cos\theta$. The construction of the spherical harmonics for non-zero values of $m$ are 
obtained from the recursion relation
\begin{equation}
P_{\ell}^{m}(u)=(-1)^m(1-u^2)^{m/2}\frac{\dd^m}{\dd u^m}P_{\ell}(u)\,.
\end{equation}
Each of these operators can then be evaluated in the flat-sky limit, with $u\approx 1-\theta^2/2$,
\begin{eqnarray}
(1-u^2)^{\frac{m}{2}}\left(\frac{-\dd}{\dd
    u}\right)^m&=&\theta^{m}\left(\frac{\dd}{\theta\dd
    \theta}\right)^m\\
&+&\frac{m(m-1)}{6}{\theta^{m}}\left(\frac{\dd}{\theta\dd \theta}\right)^{m-1}+\dots\,.\nonumber
\label{RaisemExp}
\end{eqnarray}
Taking advantage that for finite values of $m$
\begin{eqnarray}
\left[\frac{(\ell-m)!}{(\ell+m)!}\right]^{1/2}&=&\frac{1}{(\ell+1/2)^{m}}\nonumber\\
&&\hspace{-2cm}\times\left[1+\frac{m(2m+1)(2m-1)}{24\,\ell^2}+\dots\right],
\label{factorExp}
\end{eqnarray}
we can use the relation (\ref{Plexpan2}) to get the expression of $Y_{\ell}^m$, and its next to leading order corrections to an arbitrary order, in the flat-sky approximation. The final result can be written in terms of Bessel functions of the first kind thanks to the relation,
\begin{equation}
J_{m}(\theta)=(-\theta)^m\left(\frac{\phantom{\,}\dd}{\theta \dd \theta}\right)^{m}J_{0}(\theta).
\end{equation}

We then recover the known large $\ell$ expression of the spherical harmonics,
\begin{equation}
Y_{\ell}^{m}(\theta,\varphi)\to \left(\frac{2 L}{4\pi}\right)^{1/2}(-1)^{m}J_{m}[L \theta]e^{\ii m \varphi},
\end{equation}
where the first correction term is in $1/\ell^2$ or in $\theta^2$. It comes from the sub-leading terms in 
Eq.~(\ref{Plexpan2}), the sub-leading operators in (\ref{RaisemExp}) and the
subleading terms in  (\ref{factorExp}). All these corrective terms make the
flat-sky expansions controllable when $m$ is finite but clearly not when it is
of the order of $\ell$. The corrective terms are expected to involve Bessel
functions $J_{m'}$ where $m'$ differs to $m$ by at most 2. Subleading terms in
the final results of flat-sky expressions can then be obtained in principle using relations (or similar to those) given in appendix \ref{BesselOrtho}.

\bibliography{FlatSky9}

\end{document}